# Spin Dynamics in the van der Waals Ferromagnet $CrTe_2$ Engineered by Niobium Doping

*Dhan Raj Lawati,[1†] Prem Bahadur Karki,[2†] Jitender Kumar,[1†] Karishma Prasad,[3†] Mohamed A. Elekhtiar,[4] Kai Huang,[4] Bibek Tiwari,[4] Suvechhya Lamichhane,[4] Rupak Timalsina,[1] Zane Hubble,[2] Ayodimeji E. Aregbesola,[2] John Watt,[5] Sy-Hwang Liou,[4] Evgeny Y. Tsymbal,[4] Jian Wang,[3*] Kapildeb Ambal,[2*] and Abdelghani Laraoui[1,4*]*

[1] Department of Mechanical & Materials Engineering, University of Nebraska-Lincoln, Lincoln, NE 68588, United States

[2] Department of Mathematics, Statistics, and Physics, Wichita State University, Wichita, KS 67260, United States

[3] Department of Chemistry and Biochemistry, Wichita State University, Wichita, KS 67260, United States

[4] Department of Physics and Astronomy and Nebraska Center for Materials and Nanoscience, University of Nebraska-Lincoln, Lincoln, NE 68588, United States

[5] Center for Integrated Nanotechnologies, Los Alamos National Laboratory, Los Alamos, NM, 87545 United States

[*]Corresponding Authors: alaraoui2@unl.edu, kapildeb.ambal@wichita.edu, and jian.wang@wichita.edu

[†]Equal contribution

## Abstract

Understanding and controlling spin dynamics in two-dimensional (2D) van der Waals (vdW) ferromagnets is essential for their application in magnonics and hybrid quantum platforms. Here, we investigate the spin dynamics of the vdW ferromagnet 1*T*-$CrTe_2$ and demonstrate their systematic tunability via niobium (Nb) substitution in $Cr_{1-x}Nb_xTe_2$ ($x = 0 - 0.2$). Ferromagnetic resonance (FMR) spectroscopy reveals that Nb doping enables wide-band tuning of the resonance frequency from 40 GHz down to the few-GHz regime, accompanied by a moderate increase in the Gilbert damping constant from ~0.066 to ~0.14, while preserving robust room-temperature ferromagnetism. Complementary magnetometry shows a concurrent reduction of the Curie temperature and saturation magnetization with increasing Nb content. Density functional theory calculations attribute the observed spin-dynamic trends to Nb-induced modifications of magnetic anisotropy and magnetic exchange interactions. Furthermore, $CrTe_2$ flakes (~80 nm thick) exhibit lower resonance frequencies and damping than bulk crystals, consistent with thickness, surface/interface, and shape-dependent magnetic anisotropy. These results establish Nb-doped $CrTe_2$ as a tunable vdW ferromagnet with controllable spin dynamics, extending its functionality from spintronics to broadband magnonics and quantum magnonics.

## 1. Introduction

Emerging two-dimensional (2D) van der Waals (vdW) magnets[1,2] offer a flexible platform for studying and engineering magnetism beyond the limits of conventional bulk materials. Reduced dimensionality enables unprecedented control of magnetic properties through layer-by-layer engineering,[3–5] twisting,[6] doping,[7–9] and proximity effects.[10–12] Moreover, vdW magnets can be integrated into heterostructures with other vdW materials enabling electrical, mechanical, and optical control of magnetism.[2,13,14] Their weak interlayer bonding permits easy transfer onto functional substrates such as diamond or hexagonal boron nitride (hBN). These materials host optically addressable spin qubits, such as diamond nitrogen-vacancy (NV) centers and hBN defect-based spin centers, providing long coherence times and enabling strong proximity coupling for nanoscale sensing and hybrid quantum systems.[15–20]

Despite intense research activity since the initial discovery of 2D magnetism in 2017,[3,4] the spin dynamics of vdW magnets, central to magnonics, remain poorly understood or largely unexplored in many newly discovered materials. To date, spin-wave (SW)/magnon modes have been seen in only limited number of vdW magnets, such as $CrI_3$,[21,22] $CrCl_3$,[23] $FePS_3$,[24] $MnBi_2Te_4$,[25] CrSBr[26], and α-$RuCl_3$.[27] Recent electrical detection of spin pumping experiments in $Cr_2Ge_2Te_6$/Pt flakes showed GHz magnetization dynamics with a low magnetic damping ~0.4 – 1 × $10^{-3}$.[28] Establishing broadly tunable spin-dynamic properties in vdW magnets therefore remains an important open challenge for magnonics applications.[29,30]

$CrTe_2$, a member of the transition-metal dichalcogenide (TMD) family, is a particularly promising candidate owing to its robust ferromagnetism and structural versatility. It crystallizes in several structural polymorphs including 1*H*, 2*H*, $1T_d$, and 1*T* phases.[31] Notably, the 1*T* phase of $CrTe_2$ exhibits ferromagnetism with a Curie temperature ($T_C$) above room temperature in bulk crystals,[32] exfoliated thin flakes,[20,33] and epitaxially grown monolayers.[34,35] While most thick (> 10 nm) $CrTe_2$ crystals and flakes[32,33] display in-plane magnetic anisotropy, few-monolayer epitaxial films exhibit strong perpendicular magnetic anisotropy,[34,35] highlighting $CrTe_2$ as a versatile vdW ferromagnet for spintronic applications.[14,36,37] Recent multiscale modeling showed that interfacing $CrTe_2$ layer with different Te-based layers enabled engineering Dzyaloshinskii-Moriya interaction (DMI)[38], relevant to topological spin textures.[39–42] Very recently there has been a report on the exploration of ferromagnetic resonance spectroscopy of self-intercalated derivative $Cr_{1.8}Te_2$ thin films.[43] However, a systematic exploration of spin dynamics properties of vdW $CrTe_2$ and its tunability using transition metal-doped derivatives have remained largely unexplored.

Here, we explore and engineer spin dynamics in the vdW ferromagnet 1*T*-$CrTe_2$ by investigating its dynamic magnetic response using ferromagnetic resonance (FMR) spectroscopy at room temperature. To achieve systematic and controllable tunability of these spin-dynamic properties without suppressing robust room-temperature ferromagnetism, we employ niobium (Nb) substitution in 1*T*-$Cr_{1-x}Nb_xTe_2$ ($x$ = 0 – 0.2) crystals grown by direct synthesis. Nb doping provides a chemically stable and scalable route to modify magnetic interactions, thereby enabling control of spin dynamics. Structural characterization confirms high crystalline quality and Nb incorporation, while magnetometry shows that all doped crystals remain ferromagnetic, with Curie temperature decreasing modestly from ~315 to ~295 K and the saturation magnetization reduced with increasing Nb content. FMR measurements reveal that Nb doping enables wide-band tuning of spin dynamics, yielding a progressive reduction of the resonance frequency and a moderate increase of the Gilbert damping constant α from ~0.066 to ~0.14. Density functional theory (DFT) calculations attribute these trends to Nb-induced modifications of magnetic anisotropy, spin-orbit coupling, and exchange coupling. Lower (a few GHz) ferromagnetic resonance frequencies and damping (~0.04) of ~80-nm -thick $CrTe_2$ flakes compared to bulk crystals are explained by thickness, surface/interface, and shape-dependent magnetic anisotropy.[35] Together, these results establish Nb-doped $CrTe_2$ as a vdW ferromagnet with widely tunable spin dynamics, extending its applicability to magnon spintronics,[44,45] topological magnonics,[46,47] and quantum magnonics.[45,48–51]

## 2. Results and Discussion

### 2.1. Crystal Growth and Structural Characterization

1*T*-$Cr_{1-x}Nb_xTe_2$ ($x$ = 0 – 0.20) crystals were prepared via oxidation of their respective precursors $KCr_{1-x}Nb_xTe_2$ ($x$ = 0 – 0.2). The precursors were synthesized by direct solid-state reaction method[52,53] of the constituent elements (see Experimental Section and Figure S1.1, Supporting Information). X-ray diffraction (XRD) spectroscopy was performed on $Cr_{1-x}Nb_xTe_2$ ($x$ = 0, 0.05, 0.10, 0.15, 0.20), see Supporting Information S1.2 and Figure S1.2.1. The lattice parameter ($a$=$b$) of the parent pristine $CrTe_2$ crystal is 3.7803 Å and decreases to 3.7761 Å with Nb doping up to 20%. A monotonic shift of the Bragg peak towards smaller angles with increasing the Nb concentration is observed (Figure S1.2.1b, Supporting Information), indicating a lattice contraction in the *ab* plane (see Figure 6a) due to the incorporation of Nb into the $CrTe_2$ lattice. An increase in the $c/a$ ratio (Figure S1.2.1c, Supporting Information) with increasing the Nb concentration further confirms the lattice expansion along the crystallographic $c$ axis. Single-crystal X-ray diffraction (SCXRD) on 1*T*-$CrTe_2$, $Cr_{0.9}Nb_{0.1}Te_2$, and $Cr_{0.8}Nb_{0.2}Te_2$ were also employed to confirm the expansion of the $c$ axis upon Nb-doping. As shown in Figure S1.2.2 (Supporting Information), increasing Nb doping systematically expands the $c$-axis, which agrees well with powder XRD results. Such contraction/expansion of the unit cell have been observed upon replacing Cr with vanadium (V) in the first report of ferromagnetism in $CrTe_2$ crystals.[32] Changes in the lattice parameters have a distinct effect on the electrical and magnetic properties of $CrTe_2$.[32] For instance, a reduction in the magnetic moment by 60% in the room-temperature electrical conductivity was observed in 2% V-doped $CrTe_2$, which is attributed to the onset of antiferromagnetic (AFM) coupling.[32] Several studies have further indicated that the stabilization of ferromagnetic or AFM ground states in $CrTe_2$ is strongly correlated with the lattice parameters.[54,55] Theoretical calculations showed that the ground state of a monolayer is Z-type AFM, associated with an increase of the in-plane lattice constant from ~3.57 Å in the monolayer to ~3.8 Å in the bulk form.[54,55] The XRD and DFT calculation (Section 2.4) results corroborate with our bulk magnetic properties of the $Cr_{1-x}Nb_xTe_2$ crystals, discussed in detail in Section 2.2.

Selected $Cr_{1-x}Nb_xTe_2$ ($x$ = 0 – 0.2) crystals were further characterized using high-resolution transmission electron microscopy (HRTEM). Figure 1a and Figure S1.3.1 (Supporting information) show HRTEM images of different $Cr_{0.8}Nb_{0.2}Te_2$ flakes, obtained by grinding large crystals, confirming their polycrystalline nature, consistent with the XRD peaks in Figure S1.2.1a, Supporting Information. Zoomed HRTEM images of Flake 1 in Figure 1b and Flake 2 in Figure S1.3.2b show lattice spacing $d$ of ~0.22 nm and ~0.32 nm consistent with the {102} and {100} interplanar spacings, corroborating the XRD results.

To further check the presence of Nb, X-ray photoelectron spectroscopy (XPS) and energy-dispersive X-ray spectroscopy (EDS) were conducted on selected $Cr_{1-x}Nb_xTe_2$ crystals. As expected, no Nb was detected in pristine $CrTe_2$ crystals in the XPS spectra (Figure 1c). However, for $Cr_{0.8}Nb_{0.2}Te_2$ crystal (Figure 1d), there are two measured peaks (filled circles) at binding energies of ~209.58 eV and ~206.78 eV, related to Nb $3d_{3/2}$ and $3d_{5/2}$ electronic states, respectively.[56] The measured peaks in Figure 1d were fitted (solid lines) using a combined Lorentzian and Gaussian functions.[57,45] XPS peaks of Te and Cr were obtained in both pristine and $Cr_{0.8}Nb_{0.2}Te_2$ crystals (see Figure S1.4, Supporting Information) with binding energies of ~572.5 (582.8) eV for Te $3d_{5/2}$ ($3d_{3/2}$) doublets, and ~575.8 (586.4) eV, for Cr $2p_{3/2}$ ($3p_{1/2}$) doublets, respectively.[35] Although the Nb 3d peaks confirm qualitative incorporation of Nb, peak overlap of Cr and Te signals, together with the surface-sensitive nature of XPS, prevents precise quantification of the Nb concentration. Figure 1f shows EDS spectrum taken from the area (solid square) in the SEM image (Figure 1e) with Cr, Nb, Te peaks in the L and K series of $Cr_{0.8}Nb_{0.2}Te_2$ crystal. There are other peaks of O and K coming from air/oxidation and K residue during sample preparation, respectively. The EDS maps of Cr (Kα1), Nb (Lα1), and Te (Lα1) in Figures 1g-i show their homogeneous distributions over the entire region of the crystal. While our magnetic (SQUID, FMR) and XRD measurements agree well with DFT calculations on the effect of Nb doping, future synchrotron high-resolution XRD and Neutron scattering[58] measurements may reveal the exact content of Nb.

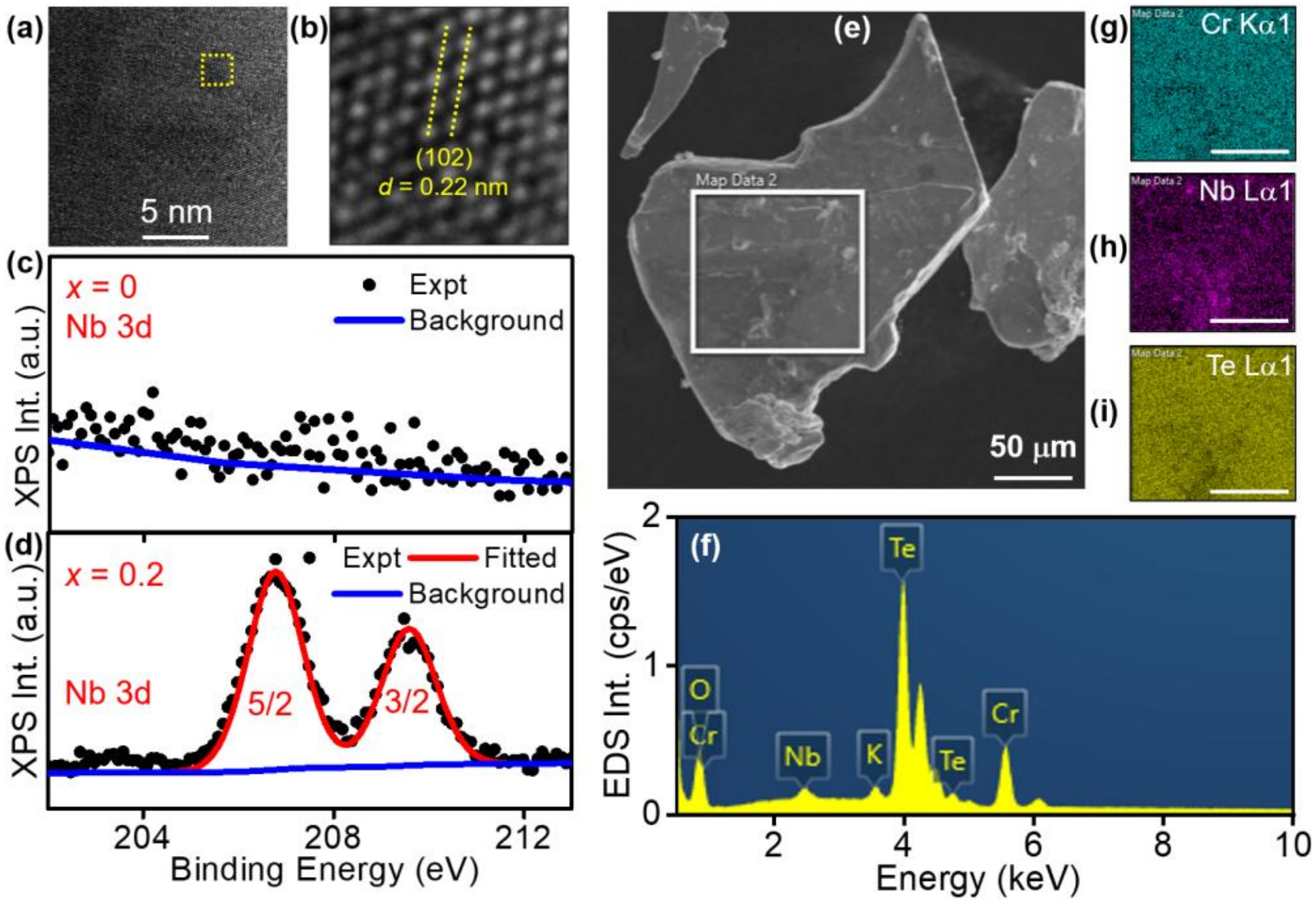

**Figure 1**. (a) HRTEM image of 1*T*-$Cr_{0.8}Nb_{0.2}Te_2$ flake, showing different crystal orientations. (b) Zoomed section of HRTEM image in (a) showing the (102) plane. High Resolution XPS spectra of $CrTe_2$ (c) and $Cr_{0.8}Nb_{0.2}Te_2$ (d) crystals. The scattered curves are measured XPS curves while the solid red lines are fits. The Blue line shows the background coming from inelastic scattered electrons. (e) SEM image of a selected $Cr_{0.8}Nb_{0.2}Te_2$ micro-crystal. (f) SEM-EDS spectrum showing the presence of Cr, Nb, Te in addition to O due to oxidation of the surface and K, a residue from the sample preparation. SEM-EDS element maps of Cr Kα1 (g), Nb Lα1 (h), and Te Lα1 (i), respectively, highlighted by a solid square in (e).

We performed additional EDS mapping using scanning transmission electron microscopy (STEM) and high-angle annular dark field (HAADF) on selected thin $Cr_{0.8}Nb_{0.2}Te_2$ flakes confirming the mixing of Nb into $CrTe_2$ crystal, see Figure S1.5 in the Supporting Information. Although, due to inherent limitations in SEM or STEM EDS, particularly in accurately measuring the content of lower doped crystals,[45] it cannot be used here to quantify the exact Nb content. Electron energy loss spectroscopy (EELS) on selected $Cr_{0.8}Nb_{0.2}Te_2$ flakes confirm the distribution of Nb within the crystal similar to Cr and Te (Figure S1.6, Supporting Information). Still, the resolution of our EELS-STEM, hindered by the thick (≥ 100 nm) nature of our flakes, is not enough to show the position of the Nb atoms near the Cr atoms. Future approaches such as exfoliating thinner (< 20 nm) flakes[33,59] or epitaxially growing thin $Cr_{1-x}Nb_xTe_2$ crystals films [34,60] may reveal the exact Nb concentration and position within the $Cr_{1-x}Nb_xTe_2$ crystals.

### 2.2. Static magnetic properties of $1T$-$Cr_{1-x}Nb_xTe_2$ ($x = 0 – 0.2$) crystals

To study the effect of Nb doping on the magnetic properties of $1T$-$CrTe_2$ crystals/flakes, we performed bulk magnetometry measurements using vibrating-sample magnetometer (VSM, PPMS Dynacool) and superconducting quantum interference device (SQUID, MPMS). Zero field cooling (ZFC) and field cooling (FC) protocols were incorporated to assess the magnetic behavior of the crystals. Figure 2a displays the ZFC $M$-$T$ curves of $Cr_{1-x}Nb_xTe_2$ ($x$ = 0, 0.05, 0.1, 0.12, 0.15, and 0.2) recorded at a magnetic field $H$ of 30 mT and show a ferromagnetic-paramagnetic phase transition in a temperature range of 315 – 295 K. To confirm the nature of magnetism in $1T$-$Cr_{1-x}Nb_xTe_2$ crystals, the magnetization $M$ is plotted versus $H$ for $x$ = 0, 0.05, 0.1, 0.12, 0.15, and 0.2 Nb-doped crystals (Figure 2b). The hysteresis loops for all crystals show a saturation behavior at $H$ ~150 mT and coercive magnetic field $H_c$ ~ 0.5 – 1.12 mT (inset of Figure 2b), confirming the ferromagnetic nature of the pristine $1T$-$CrTe_2$[32,35] and Nb-doped crystals.

To examine the magnetic behavior in detail, we performed magnetic measurements on $Cr_{1-x}Nb_xTe_2$ at different times and with varying sizes and shapes (crystals, nanoflakes). See Section S2 in the Supporting Information. The values of the Curie temperature ($T_C$) and saturation magnetization ($M_S$) presented in Figure 2 are obtained by averaging results from multiple measurement runs in Section S2, Supporting Information, except for 12% Nb concentration. $T_C$ is retrieved from the derivatives of $M$-$T$ curves in the inset of Figure 2a and is plotted as a function of Nb-doping factor $x$ in Figure 2c. $T_C$ decreases from 314.79 ± 0.22 K for pristine crystals to 295.04± 0.61 K for the $Cr_{0.8}Nb_{0.2}Te_2$ crystals. A similar trend for $M_S$ vs Nb doping is observed, i.e., $M_S$ decreases from 77.19 ± 1.19 kA/m for pristine crystals to 31.58 ± 0.29 kA/m for $Cr_{0.8}Nb_{0.2}Te_2$ crystals.

We observe a noticeable variation in $T_C$ and $M_S$ for crystals with higher Nb doping levels (e.g., 10% and 20%). This behavior may be attributed to the metastable nature of $1T$-$CrTe_2$ doped compounds and their susceptibility to oxidation,[32] in particular for the 20% Nb doped crystals with a significant reduction of $M_S$ (~7.86 kA/m) and $T_C$ (~284 K) with repeated magnetic measurements after 385 days (see Figure S2.5, Supporting Information). As for the 10% Nb doped crystals, there is a variation of $T_C$ from ~309 to 316 K, indicative of a variation of the exact Nb during the initial growth.

The systematic reduction in $T_C$ and $M_S$ with increasing Nb doping, despite the contraction of the lattice along the *ab* plane (see Figure 6a), can be understood in terms of magnetic-site dilution and changes in exchange interactions. With increasing Nb content, non-magnetic Nb ions replace Cr sites, weakening the overall ferromagnetic coupling. In addition, the contraction of the *ab*-plane enhances the Cr–Cr direct exchange, which favors antiferromagnetic interactions, while the Cr–Te–Cr super-exchange that supports ferromagnetism becomes relatively less dominant. Simultaneously, the lattice expansion along the *c*-axis reduces the magnetic anisotropy energy, further lowering the Curie temperature, further discussed in Section S.2 and Figure S2.6 in the Supporting Information.

In-plane (IP) and out-of-plane (OOP) SQUID measurements on bulk pristine $1T$-$CrTe_2$ crystals (see Figure S2.7a, Supporting Information) confirm the in-plane magnetic anisotropy with a low saturation field ~150 mT for IP in comparison to > 300 mT for OOP measurements. Due to the complex geometry of the crystals, it is not possible to deduce the amplitude of magnetic anisotropy as it is done in $1T$-$CrTe_2$ films.[34,35] We used FMR to deduce the effective magnetic anisotropy ($K_{eff}$) of pristine and Nb doped crystals, as discussed below.

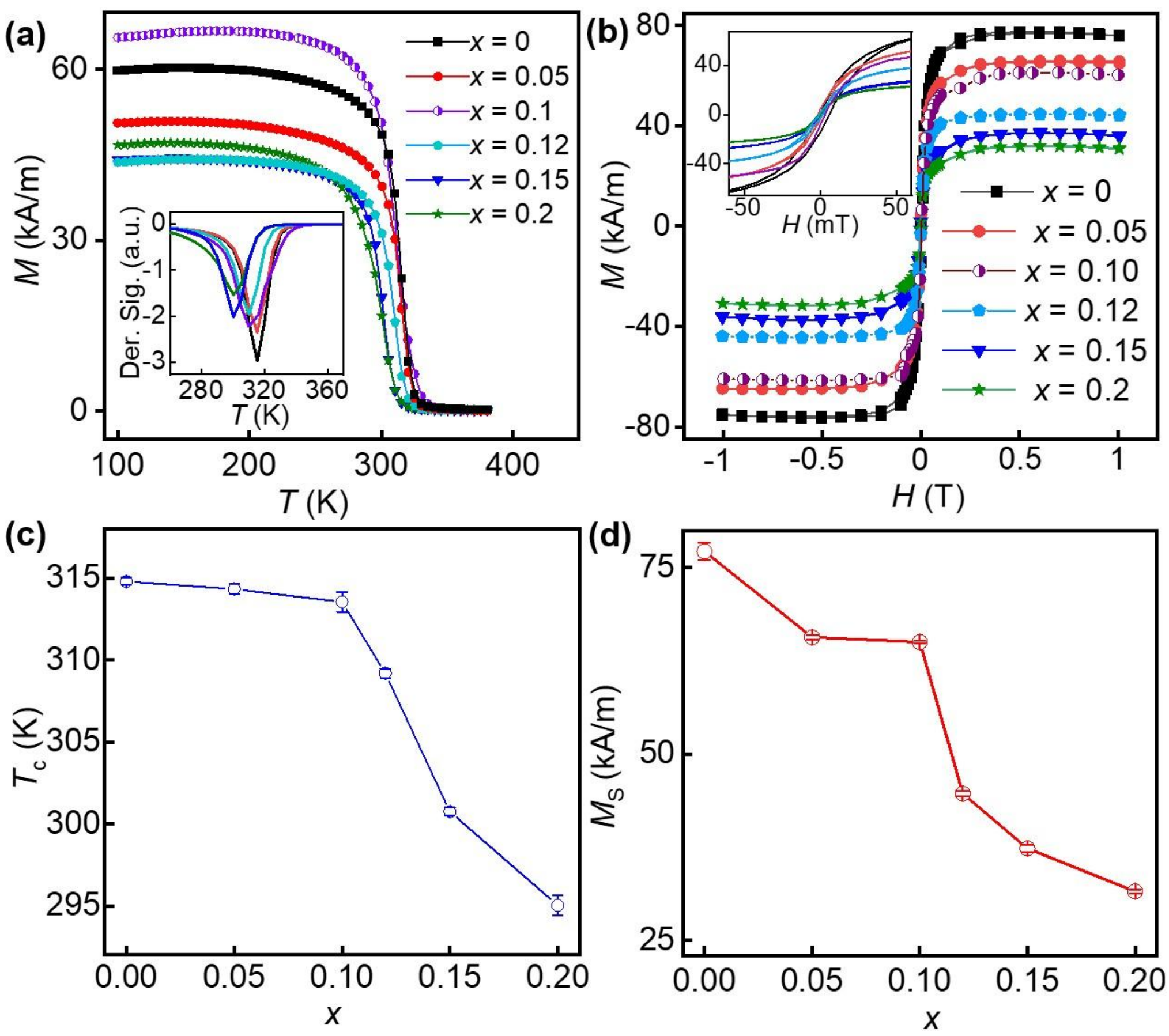

**Figure 2**: Magnetic measurements of 1*T*-$Cr_{1-x}Nb_xTe_2$ ($x$ = 0, 0.05, 0.10, 0.12, 0.15, 0.2) crystals. (a) ZFC IP magnetization ($M$) versus temperature (T) curves at a magnetic field of 30 mTof $Cr_{1-x}Nb_xTe_2$ crystals, showing a ferromagnetic to paramagnetic phase transition in the range of 315– 295K. Inset of (a): derivatives of M-H curves in (a) to extract $T_C$. (b) IP *M-H* hysteresis loops measured at 300 K of $Cr_{1-x}Nb_xTe_2$ crystals. Inset of (b): zoomed *M-H* loops showing a coercive field in the range ~0.5 – 1.12 mT. $T_C$ (c) and $M_S$ (d) as a function of Nb doping concentration factor $x$.

## 2.3. Spin dynamics of 1*T*-$Cr_{1-x}Nb_xTe_2$ ($x$ = 0 – 0.2) crystals and flakes

To study the spin dynamics of 1*T*-$Cr_{1-x}Nb_xTe_2$ crystals and flakes, we performed FMR spectroscopy in a microwave (MW) frequency range of 1 – 40 GHz as a function of the applied magnetic field $H$ (up to 0.6 T). Figures 3a and 3b show the schematic and picture of the coplanar waveguide (CPW) FMR setup, respectively, described in the Experimental Section. In Figures 3(c-g), we plot the normalized FMR (derivative MW absorption) signal versus the MW frequency and applied magnetic field for $Cr_{1-x}Nb_xTe_2$ ($x$ = 0, 0.05, 0.1, 0.15, and 0.2) crystals, done at room temperature (~297 K).

The spin dynamics of the 1*T*-$Cr_{1-x}Nb_xTe_2$ crystals can be expressed by the Landau-Lifshitz-Gilbert (LLG) equation:[61]

$$\frac{\partial \boldsymbol{M}}{\partial t} = -\gamma \boldsymbol{M} \times \boldsymbol{H}_{eff} + \alpha \, \frac{\boldsymbol{M}}{M_S} \times \frac{\partial \boldsymbol{M}}{\partial t} \, ,$$

where $\boldsymbol{H}_{\text{eff}}$ is the effective magnetic field that contains both the applied magnetic field $H$ and anisotropy/exchange fields, $\boldsymbol{M}$ is the magnetization vector, $\alpha$ is the Gilbert damping constant. $\gamma$ is the gyromagnetic ratio, defined as $\gamma = -g \, {}^{\mu_B}/_{\hbar}$ ($g$ is the Landé factor and $\mu_B$ is the Bohr magneton).

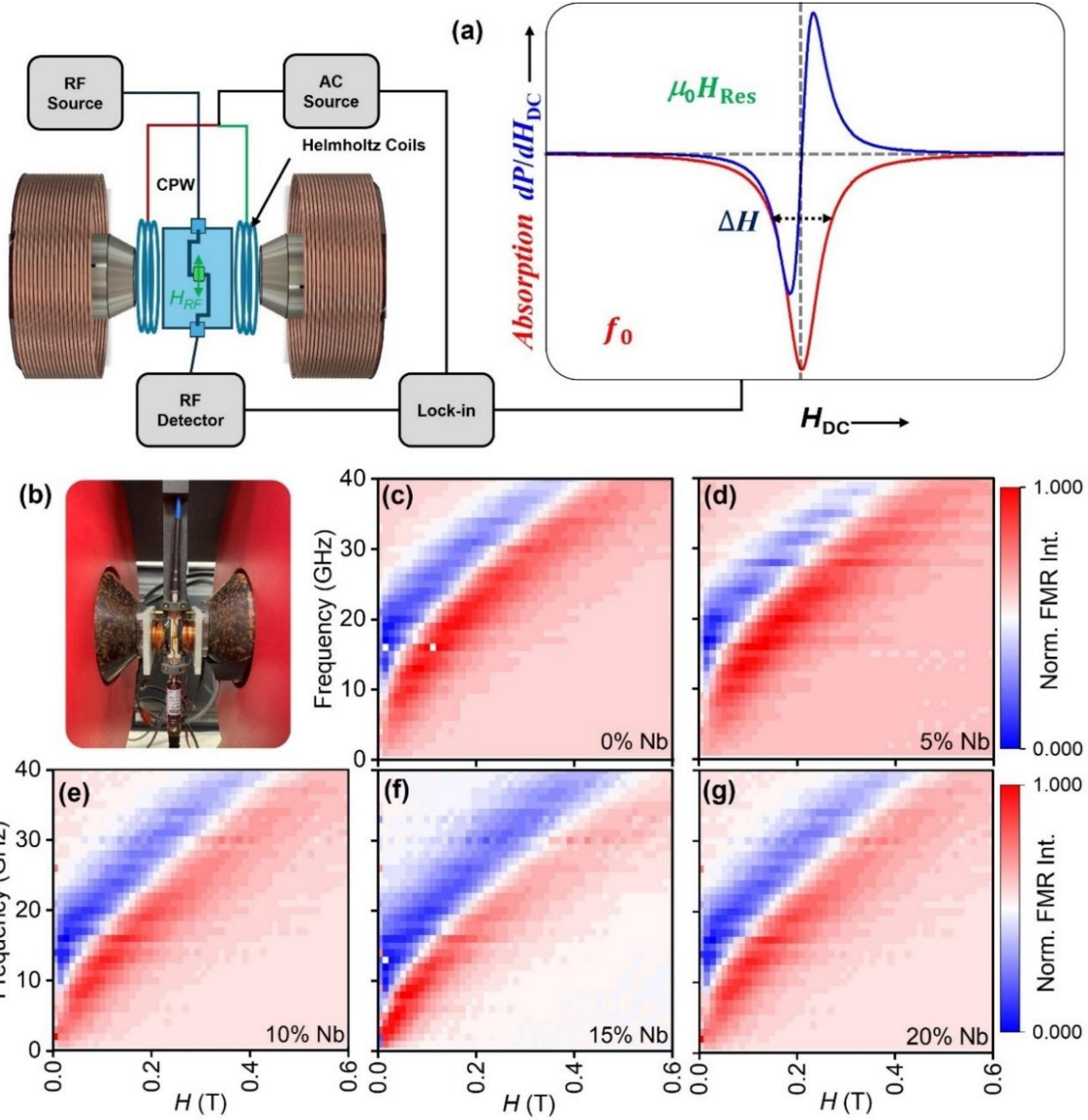

**Figure 3.** (a) Schematic representation of the FMR setup with diagram of MW absorption and lock-in detected derivative spectra. (b) A picture of the FMR setup: the sample is placed on top of a 50 Ω-matched CPW with one end connected to the MW source and the other to the MW detector. The CPW is placed between the poles of Helmholtz coils electromagnet. (c-g) color plots of the normalized FMR derivative absorption signal for 1$T$-$Cr_{1-x}Nb_xTe_2$, $x$ = 0, 0.05, 0.1, 0.15, and 0.2, respectively.

To deduce the Nb doping effect on the magnetic anisotropy, damping, and Landé $g$-factor, the frequency dependence of the ferromagnetic resonance field $H_R$ is plotted in Figure 4a and fitted (solid line) with the Kittel formula for an in-plane sample as:[61,62]

$$f = \left(\mu_0\gamma/2\pi\right)\sqrt{H_R(H_R + M_{eff})},$$

where $\mu_0$ is the vacuum magnetic permeability and $M_{eff} = M_s - H_k$.[62] $H_k$ is the effective magnetic anisotropy field containing both surface/interface anisotropy and exchange fields. For bulk crystals $H_k = 2K_{eff}/\mu_0 M_S$. $K_{eff}$ is ~1.47 ± 0.02 × $10^5$ J/m$^3$ for pristine 1$T$-$CrTe_2$ crystals which is more than the measured value at 300 K of 0.49 × $10^5$ J/m$^3$ in ultrathin (thickness of 7ML) epitaxial 1*T*-$CrTe_2$ films.[34]

The Nb doping strongly affects $K_{eff}$ as shown in Figure 4b. $K_{eff}$ decreases from 1.47 ± 0.02 × $10^5$ J/m$^3$ for the parent pristine $CrTe_2$ crystals to ~0.34 ± 0.02 ×$10^5$ J/m$^3$ for the $Cr_{0.8}Nb_{0.2}Te_2$ crystals, which is explained by the decrease in magnetic exchange coupling between Cr atoms upon Nb (bigger atoms) doping. Similar behavior is observed in the DFT calculations at a temperature of 297 K, discussed in Section 2.4. The Landé $g$-factor increases with increasing the Nb doping concentration as: $g$ = 2.00924 for the pristine $CrTe_2$ and 2.125 for the $Cr_{0.8}Nb_{0.2}Te_2$ crystals. Similar effects were observed in the V-doped $Fe_{1-x}V_x$ alloys.[62]

We performed IP and OOP FMR measurements on pristine 1*T*-$CrTe_2$ crystals (Figure S2.7b, Supporting information). The higher resonance field $H_R$ for the OOP configuration confirms in-plane magnetic anisotropy.[42,45]

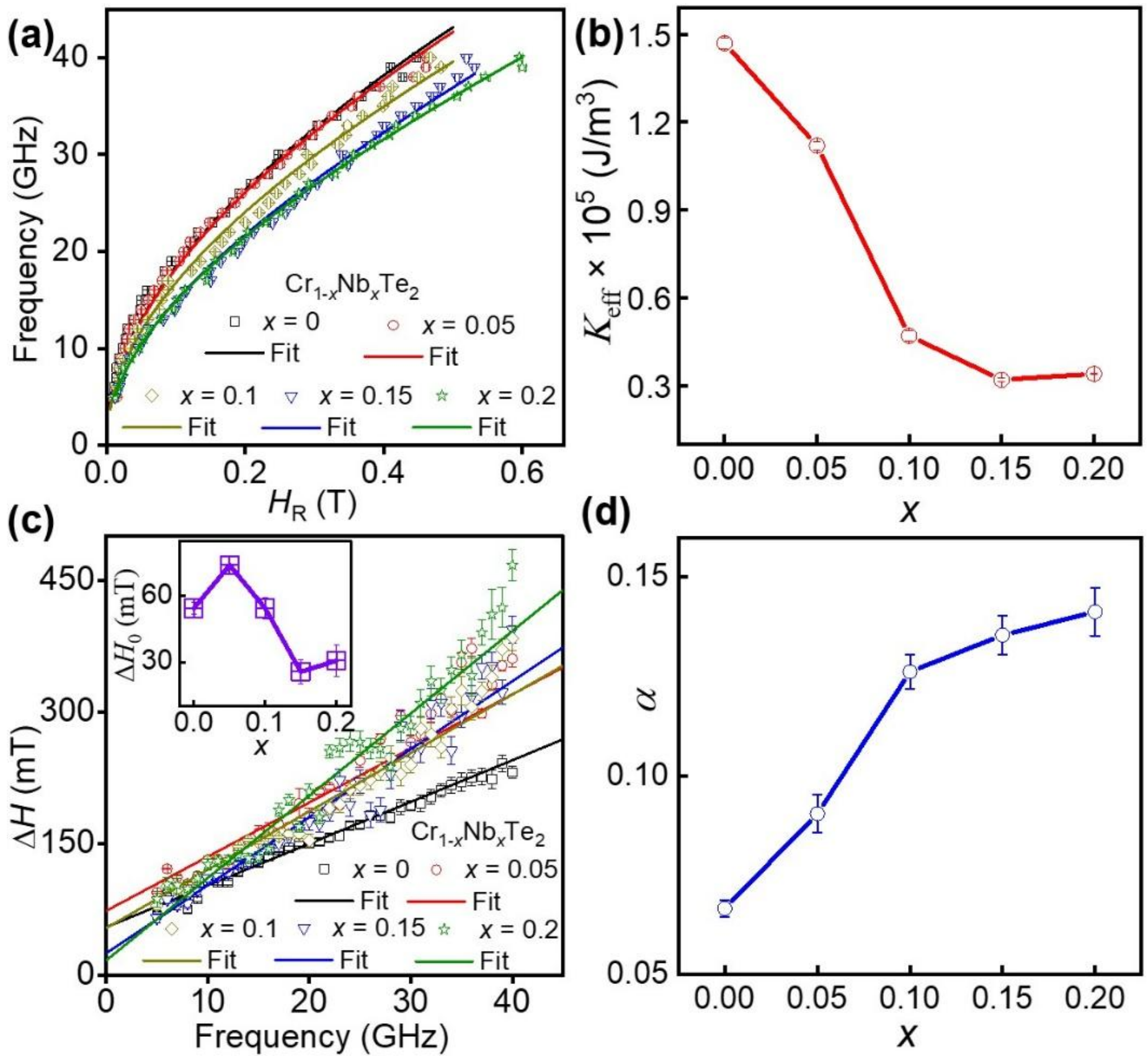

**Figure 4.** (a) Frequency dependence of the resonance field vs applied magnetic field for $Cr_{1-x}Nb_xTe_2$, $x$ = 0, 0.05, 0.1, 0.15, and 0.2, respectively. The solid lines in (a) are fit using the Kittel formula for in-plane applied magnetic field. (b) $K_{eff}$ versus Nb doping factor $x$. (c) Measured (scattered) FMR linewidth extracted from the plots in Figures 3(c-g), fitted (solid lines) with linear fits. Inset of (c) is the intercept $\Delta H_0$. (d) $\alpha$ vs Nb doping factor $x$.

To extract $\alpha$, the FMR linewidth $\Delta H$ is plotted vs the MW frequency $f$ in Figure 4c, and linear fitted (solid lines) as:[63]

$$\Delta H = 4\pi\alpha f/\gamma\mu_0 + \Delta H_0,$$

where $\Delta H_0$ is found in the range of 25 – 74 mT (see inset of Figure 4c). As shown in Figure 4d, $\alpha$ increases from 0.0666 ± 0.0021 for pristine $CrTe_2$ to 0.1413 ± 0.0060 for $Cr_{0.8}Nb_{0.2}Te_2$ crystals, respectively. Nb increases the damping due to the large size (0.68 Å) of the Nb atoms in comparison to the Cr (0.55 Å) atoms that enhances spin-orbit coupling with Te layer and scattering of conduction electrons.[31,32] The damping of the pristine and Nb doped $CrTe_2$ is comparable to ferromagnetic metallic (e.g., Co and $CoPt_3$),[64,65] ferrimagnetic insulating (e.g., TmIG)[45,66] films, and recently measured $Cr_{1.8}Te_2$ epitaxial films.[43]

We performed FMR spectroscopy of pristine $CrTe_2$ flakes with an average thickness of ~80 nm, as determined from atomic force microcopy (Section S3 and Figure S3.1 in the Supporting Information). The $CrTe_2$ flakes (Figure 5a) were exfoliated from the pristine 1*T*-$CrTe_2$ crystals using vinyl blue tape[67,68] and placed on top of sapphire substate with MW antenna. See supporting Information Section S3 for further details about the FMR measurements. Figure 5b shows the derivative of the measured normalized FMR absorption spectrum vs the external magnetic field $H$ at a MW frequency of 17 GHz. The FMR curves at each MW frequency were fitted (solid lines in Figure 5b) to extract the resonance field $H_R$. The frequency dependence of $H_R$ is plotted in Figure 5c and fitted with in-plane Kittel formula,[61] giving an effective magnetization $M_{eff}$ of ~1.13 ± 0.3 T. Due to the low volume of the magnetic flakes, it is not possible to use standard bulk magnetometry techniques (e.g., SQUID) to deduce $M_S$ of the $CrTe_2$ flakes. In reference [33], NV magnetometry was used to deduce $M_S$ ~27 kA/m for thin (thickness of ~20 nm) $CrTe_2$ flake from magnetic stray-field measurements.[33] This value is way less than the measured value of ~77 kA/m of the pristine $CrTe_2$ crystals. By assuming an $M_S$ of the 80-nm flakes is ~77 kA/m, we deduce an effective magnetic anisotropy $K_{eff}$ of ~$0.4 \times 10^5$ J/m$^3$, which is close to the value of ~$0.49 \times 10^5$ J/m$^3$ obtained at 300 K in 7 ML $CrTe_2$ films.[34] The $CrTe_2$ flake's magnetic anisotropy is less than the bulk (crystal) value of ~$1.47 \times 10^5$ J/m$^3$, explained by magnetic anisotropy thickness, surface/interface, and shape dependence effects.[34,35,69]

To extract the damping constant $\alpha$ of the $CrTe_2$ flakes, $\Delta H$ is plotted versus the frequency $f$ in the inset of Figure 5c and linear fitted. $\alpha$ is 0.0417, less than the bulk value of ~0.0665, that is explained by surface (roughness, interface) effects[70] and comparable to the measured value for thin $Cr_{1.8}Te_2$ films.[43]

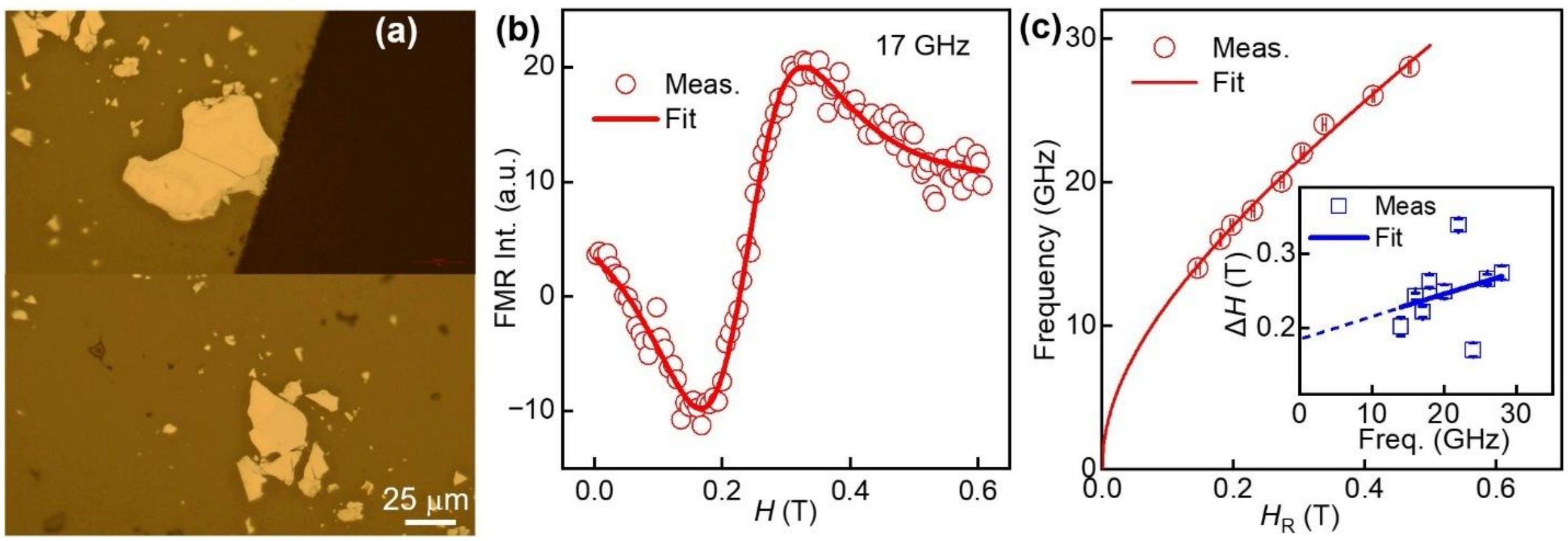

**Figure 5.** (a) Optical image of 1*T*-$CrTe_2$ flakes, exfoliated from bulk 1*T*-$CrTe_2$ crystal and transferred to the CPW antenna for FMR measurements. (b) FMR intensity vs applied magnetic field at a MW frequency of 17 GHz of $CrTe_2$ flakes in (a). (c) Measured (scattered) and fitted (solid line) frequency dependence of the resonance field $H_R$ vs applied magnetic field. Inset of (c): Measured (scattered) FMR linewidth $\Delta H$ extracted from the plot in (c), fitted (solid lines) with linear fit to extract $\alpha$.

### 2.4. Density Functional Theory Calculations

To study the Nb doping effect on the magnetic properties of 1$T$-$Cr_{1-x}Nb_xTe_2$ crystals, electronic structure DFT calculations were performed at 0 K temperature using the plane-wave pseudopotential method, incorporated in *Quantum ESPRESSO*.[71] Fully relativistic optimized norm-conserving pseudopotentials[72] were employed, and the exchange–correlation functional was analyzed using the generalized gradient approximation (GGA).[73] Magnetic effects were accounted by including spin–orbit coupling in the pseudopotentials. The virtual crystal approximation (VCA)[74] was applied to model the Nb doping, by implementing a kinetic energy cutoff of 100 Ry for the plane-wave basis set, and Gaussian smearing with a broadening of 0.01 Ry was adopted. Brillouin-zone integrations were made on a 24 × 24 × 15 Monkhorst–Pack $k$-point mesh to ensure convergence for all bulk calculations. For each Nb doping concentration factor $x$, the primitive unit cell structure was fully relaxed, and the corresponding lattice constants are shown in Figure 6c. The Hamiltonian of the system is expressed as:

$$H = -\sum_{i\neq j} J_{\parallel} \boldsymbol{s}_i \cdot \boldsymbol{s}_j - \sum_{i\neq j} J_{\perp} \boldsymbol{s}_i \cdot \boldsymbol{s}_j - K \sum_{i} (\hat{\boldsymbol{z}} \cdot \boldsymbol{s}_i)^2,$$

where $\boldsymbol{s}$ denotes the normalized spin vector, $J_{\parallel}$ is the exchange coupling between in-plane nearest neighbors (Figure 6a), $J_{\perp}$ is the exchange coupling between adjacent layers (Figure 6b), and $K$ is the magnetic anisotropy energy per magnetic atom. The exchange interaction $J_{\parallel}$ and $J_{\perp}$ were calculated using supercells of size 2 × 2 × 1 (Figure 6a) and 1 × 1 × 2 (Figure 6b), respectively, by evaluating the energy difference among ferromagnetic and antiferromagnetic configurations, as shown in Figure 6d.

The effective magnetic anisotropy $K_{eff}$ (Figure 6f) is obtained from the total energy difference between IP and OOP ferromagnetic configurations, as detailed in the Supporting Information Section S4. The calculated values of $K_{eff}$ at 0 K are approximately one order of amplitude higher than the measured values, which can be attributed to the temperature dependence of magnetic anisotropy. According to the Callen–Callen relation,[75] $K(T)/K(0) \approx \left(1 - \frac{T}{T_c}\right)^{3\beta}$, where $\beta$ denotes the critical exponent and is typically around 0.33 for bulk 3D magnets. The anisotropy decreases with the reduction of magnetization at finite temperature. Using this relation, the estimated $K_{eff}$ at 297 K (see Section S4, Supporting Information) is close with the experimental results. With increasing Nb doping, both the magnetic moment (Figure 6e) and the magnetic anisotropy (Figure 6f) are reduced. This trend can be understood from the modification of transition-metal $d$ orbital states near the Fermi level (see Section S4, Supporting Information), which affects the spin orbit coupling and consequently the anisotropy energy. The calculated results are consistent with experimental observations.

The observed trend also correlates with theoretical calculations of the exchange parameters (Figure 6d), showing a monotonic decrease in $J_{\parallel}$ and an increase in $J_{\perp}$. The reduction of $J_{\parallel}$ can be understood by the fact that a relatively small lattice parameter $a$ favors direct antiferromagnetic Cr−Cr exchange over the indirect ferromagnetic Cr-Te-Cr exchange.[32] Conversely, the increase of $J_{\perp}$ arises from reduction of magnetic anisotropy by substituting $Cr^{4+}$ (0.55 Å) with the larger Nb ion (0.68 Å), which enhances spin–orbit coupling with the Te layer and support the ferromagnetic character of the systems. These findings highlight the tunability of $CrTe_2$'s magnetic properties, with Nb doping enabling modification of magnetic interactions while maintaining robust ferromagnetic order at room temperature.

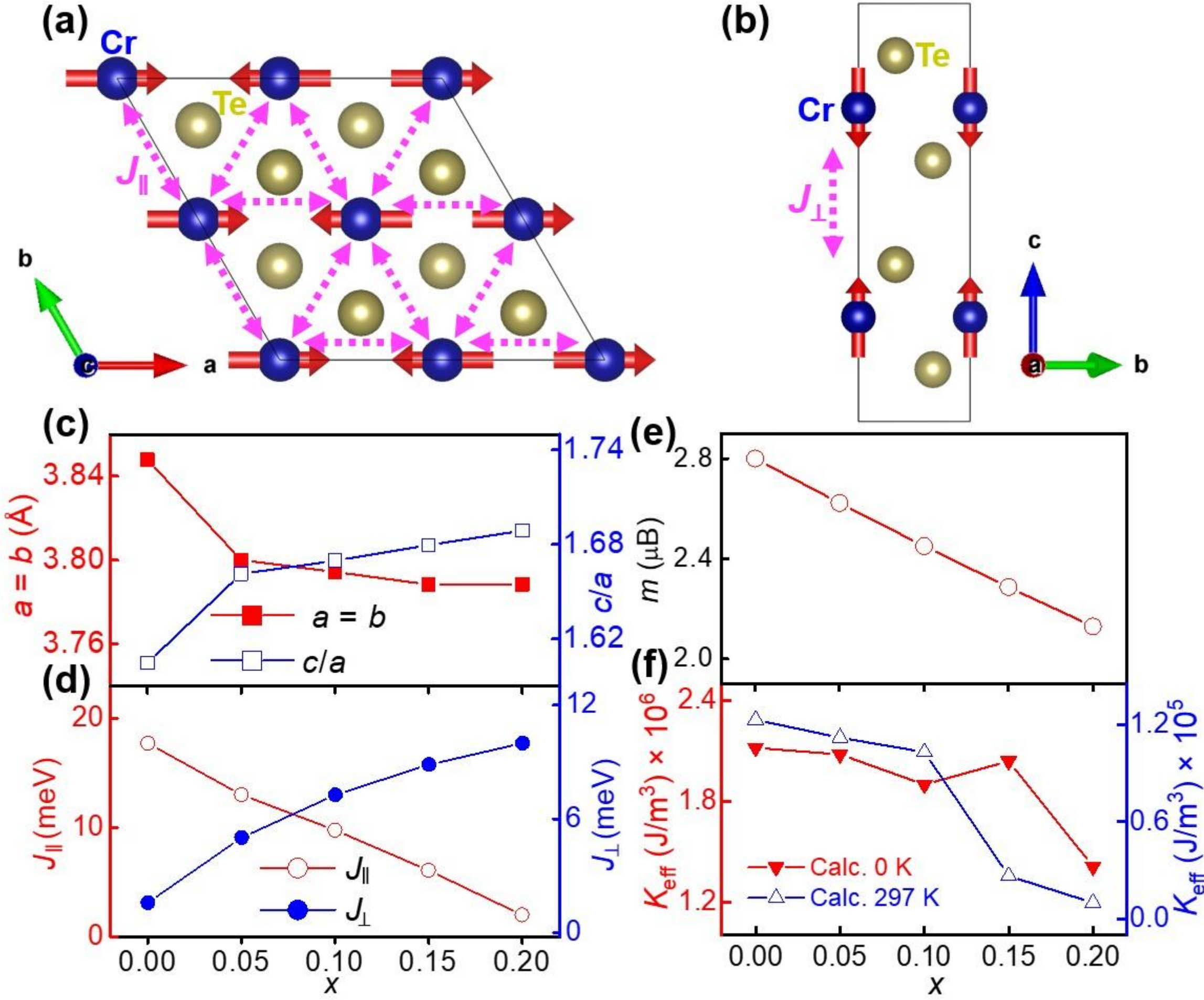

**Figure 6.** Top view (a) and side view (b) of $Cr_{1-x}Nb_xTe_2$ exchange interactions $J_{\parallel}$ and $J_{\perp}$ between neighboring magnetic atoms, indicated by the magenta arrows. (c-f) Evolution of parameters as a function of Nb concentration: (c) lattice constant $a = b$ and $c/a$, (d) exchange interactions $J_{\parallel}$ and $J_{\perp}$, (e) magnetic moment $m$, and (f) effective magnetic anisotropy $K_{eff}$ at 0 K and 297 K.

## 3. Conclusions

In conclusion, we have demonstrated that the spin dynamics of 1*T*-$CrTe_2$ crystals can be systematically engineered by niobium substitution in $Cr_{1-x}Nb_xTe_2$ ($x = 0 – 0.2$), while preserving robust room-temperature ferromagnetism. Nb doping provides an effective route to modify magnetic interactions, leading to controlled changes in both static and dynamic magnetic properties. With increasing Nb concentration, the Curie temperature $T_C$, saturation magnetization, and magnetic anisotropy are reduced, while robust ferromagnetism persists close to room temperature ($T_C$ ~295 K for the $Cr_{0.8}Nb_{0.2}Te_2$). Structural characterization using XRD and HRTEM reveals a Nb-induced lattice expansion along the *c* axis without degradation of crystallinity quality. DFT calculations reproduce these trends and attributed them to a reduction of both intralayer and interlayer exchange interactions between Cr atoms arising from a larger atomic radius of Nb compared to Cr.

Most importantly, Nb doping strongly tailors the dynamic magnetic response. FMR measurements reveal wide-band tunability of the resonance frequencies over 5 – 40 GHz, accompanied by only a modest increase of the Gilbert damping constant $\alpha$ to 0.066 – 0.14, remaining comparable to values reported for ferromagnetic metals[64,65] and ferrimagnetic insulators.[45,66] These characteristics establish Nb-doped 1*T*-$CrTe_2$ as a promising vdW ferromagnet for magnonic applications, extending its functionality beyond conventional spintronics[34] to magnon spintronics[44] and topological magnonics.[46,47]

FMR measurements on 1*T*-$CrTe_2$ flakes revealed reduced resonant frequencies down to ~2 GHz at low applied magnetic field (~3 mT), indicating additional tunability of the dynamic magnetic properties through thickness control. Although achieving homogeneous exfoliation of $CrTe_2$ flakes[33] remains challenging, controlled exfoliation assisted by thin $Al_2O_3$ capping layers[76,77] may enable the fabrication of large-area ultrathin $Cr_{1-x}Nb_xTe_2$ flakes. Such flakes provide a pathway to exploit $Cr_{1-x}Nb_xTe_2$ as a 2D vdW magnetic platform for magnonics.[37,78] Finally, integration with diamond[16–18] and/or hBN[19,20] substrates hosting spin qubits could enable proximity-driven spatial mapping of their spin-wave modes, opening avenues toward quantum magnonics and hybrid quantum architectures.[48–51,66]

## 4. Experimental Section

*Samples preparation*: A stoichiometric mixture of high-purity potassium, chromium, niobium, and tellurium was prepared according to the desired Nb doping level ($x$) under argon flow in a glove box. The element (K, Cr, Nb, Te) mixture was loaded into evacuated quartz tubes to avoid oxidation and sealed under vacuum. The sealed tubes were heated at 900 °C in 24 hours while maintaining this temperature for eight days to ensure complete reaction. After the heat treatment, the tubes were allowed to cool slowly to room temperature at a rate of 80 °C/hour. After opening the tubes in the glove box, the potassium (K) atoms were removed via a deintercalation process.[32] The resulting products consisted of dark, metallic-luster platelets and fine powders, which were handled in the glove box to prevent oxidation, Figure S1 (Supporting Information).

*Ferromagnetic Resonance spectroscopy*: Broadband Ferromagnetic Resonance (FMR) spectroscopy was performed on 1*T*-$Cr_{1-x}Nb_xTe_2$ ($x$ = 0 – 0.2) crystals and flakes by measuring the change in microwave (MW) power absorption at various MW frequencies (1 – 40 GHz) as a function of external magnetic field (see Figure 3). The samples were mounted on a copper Coplanar Waveguide (CPW) (center conductor width of 0.4 mm and a gap of 0.2 mm), oriented in such that the microwave AC magnetic field is perpendicular to the applied dc magnetic field. The signal generator generates MW frequencies ranging from 1 to 40 GHz. Electromagnets apply the static magnetic field and are continuously monitored using a Hall probe. As the magnetic field was swept through the resonance, the magnetization vector processed at resonance, it absorbed the MW energy, resulting in a decrease in the transmitted power. The power change was detected and converted into a DC voltage by a high-frequency MW detector. To improve the signal-to-noise ratio, a Lock-in Amplifier-based detection was employed. The applied magnetic field was modulated at 77 Hz with a 5 mT amplitude using a pair of Helmholtz coils. The recorded spectra represent the derivative of the RF absorption power. All the recoded data were background-subtracted, numerically integrated, and fitted with a Lorentzian function to deduce the resonance magnetic field and the FMR linewidth.

## Acknowledgements

K.A., J.W., and A.L. acknowledge the National Science Foundation (NSF) through Award 2328822. A.L. thanks for additional support from NSF Award 2521415. A.L. and E.Y.T acknowledge the University of Nebraska-Lincoln (UNL) Grand Challenges catalyst award entitled "Quantum Approaches addressing Global Threats". Theoretical modeling at UNL was supported by NSF through the EPSCoR RII Track-1 program under Award OIA-2044049 (M.A.E., K.H., E.Y.T.) and the U.S. Department of Energy, Office of Science, Office of Basic Energy Sciences, through the DOE EPSCoR program under Award No. DE-SCSC0026103 (K.H., E.Y.T.). The research performed at UNL, part in the Nebraska Nanoscale Facility and/or NERCF, was supported by NSF Award 2025298. The DFT calculations were performed at the University of Nebraska Holland Computing Center. STEM and EELS measurements were performed at the Center for Integrated Nanotechnologies, an Office of Science User Facility operated for the U.S. Department of Energy (DOE) Office of Science. Los Alamos National Laboratory, an affirmative action equal opportunity employer, is managed by Triad National Security, LLC for the U.S. Department of Energy's NNSA, under contract 89233218CNA000001.

## Conflict of Interest

The authors declare no conflict of interest.

## Author Contributions

A.L., K.A. and J.W. conceived the concept, designed the experiments, and supervised the project. D.R.L performed the magnetic (SQUID, VSM) measurements, with assistance from J.K., R.T., and A.L. K.P., Z.H., and J.W. synthesized the $CrNbTe_2$ crystals and performed XRD measurements. J.K. analyzed the XRD data and fitted the curves. P.B.K., A.E.A., and K.A performed FMR measurements. M.A.E., K.H., and E.Y.T. conducted DFT calculations. B.T. performed SEM, EDS, and HRTEM. S.L. and S.-H.L performed XPS. J.W. performed EELS on selected $Cr_{1-x}Nb_xTe_2$ crystals/flakes. A.L. wrote the manuscript with contributions and editing from all authors.

## Data Availability Statement

The data that support the findings of this study are available from the corresponding author upon reasonable request.

# Supporting Information

## Section 1. Synthesis and characterization of 1*T*-$Cr_{1-x}Nb_xTe_2$ crystals

### S1.1 Growth of $Cr_{1-x}Nb_xTe_2$ ($x$ = 0.05 – 0.2) crystals

1*T*-$CrTe_2$ and 1*T*-$Cr_{1-x}Nb_xTe_2$ ($x$ = 0.05 – 0.2) crystals were made by using direct solid-state reaction method[52] of the constituent elements (K, Cr, Nb, and Te), detailed in Figure S1.1 and the main text.

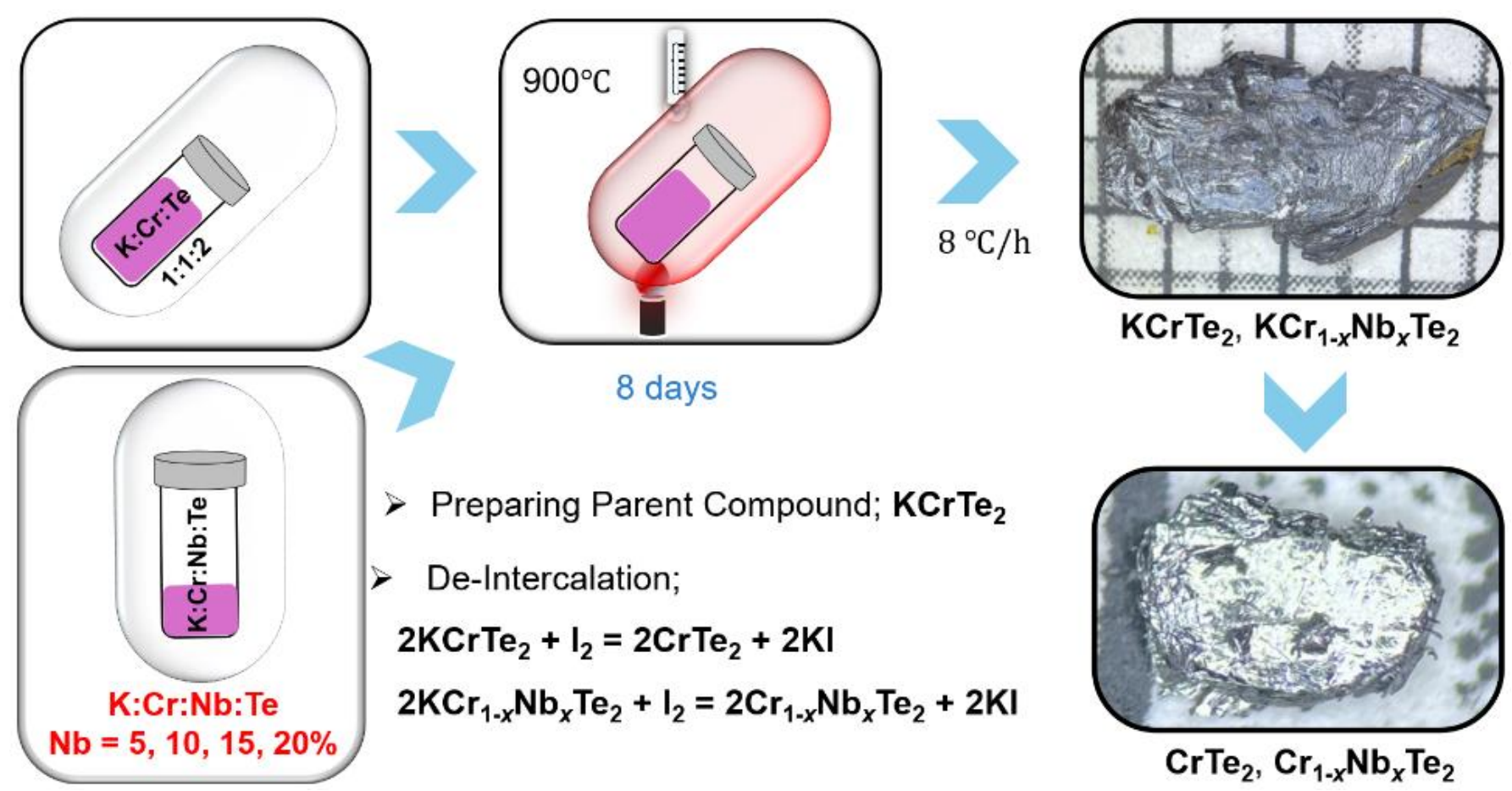

**Figure S1.1.** Synthesis of $CrTe_2$ and $Cr_{1-x}Nb_xTe_2$ ($x$ = 0.05 – 0.2) crystals using solid-state reaction method.

### S1.2 X ray diffraction of $Cr_{1-x}Nb_xTe_2$ ($x$ = 0.05 – 0.2) crystals

We performed X-ray diffraction (XRD) spectroscopy on all $Cr_{1-x}Nb_xTe_2$ crystals using Rigaku Mini Flex 6G diffractometer ($\lambda$ = 1.5406 Å). Figure S.1.2 depicts the structural information of the compounds 1*T*-$Cr_{1-x}Nb_xTe_2$ ($x$ = 0, 0.05, 0.10, 0.15, 0.20). The lattice parameters ($a$, $b$, and $c$) of the crystals in Table S1 were obtained by analysis of the diffraction pattern (2θ = 10° to 60°) by the Le Bail method using the FullProf Suite program.[79] The XRD peaks of all the compositions in Figure S1.2a are well fitted with the space group $P\bar{3}m1$, No. 164, suggesting single-phase creation of the synthesized crystals. In Figure S1.2b, a shift of the Bragg peak to smaller angles with increasing the Nb concentration is observed, indicating lattice expansion due to the incorporation of larger Nb atoms to replace Cr atoms in the lattice. Figure S1.2c shows a decrease of the lattice parameter ($a = b$) from ~3.7803 Å to ~3.7761 Å with increasing the Nb doping up to 20%. A successive increase in the $c/a$ ratio (Figure S1.2c) with higher Nb concentration further confirms the lattice expansion along the crystallographic $c$-axis. The deduced density of the $Cr_{1-x}Nb_xTe_2$ ($x$ = 0 – 0.2) crystals, plotted in Figure S1.2d, decreases with the decrease in the Nb concentration, explained by substituting the lighter Cr atoms with the heavier Nb atoms.

**Table S1:** Lattice parameters obtained from X-ray profile fitting

| Crystal composition | Lattice parameters ($a$=$b$) (Å), $\alpha$=$\beta$=90° and $\gamma$=120° | Lattice parameter $c$ (Å) | $c/a$ ratio |
|---|---|---|---|
| $CrTe_2$ | 3.7803(1) | 6.0936(2) | 1.6119(0) |
| $Cr_{0.95}Nb_{0.05}Te_2$ | 3.7808(2) | 6.0893(3) | 1.6105(1) |
| $Cr_{0.90}Nb_{0.10}Te_2$ | 3.7778(1) | 6.1031(3) | 1.6155(1) |
| $Cr_{0.85}Nb_{0.15}Te_2$ | 3.7785(1) | 6.1210(3) | 1.6199(1) |
| $Cr_{0.80}Nb_{0.20}Te_2$ | 3.7761(2) | 6.1428(5) | 1.6267(1) |

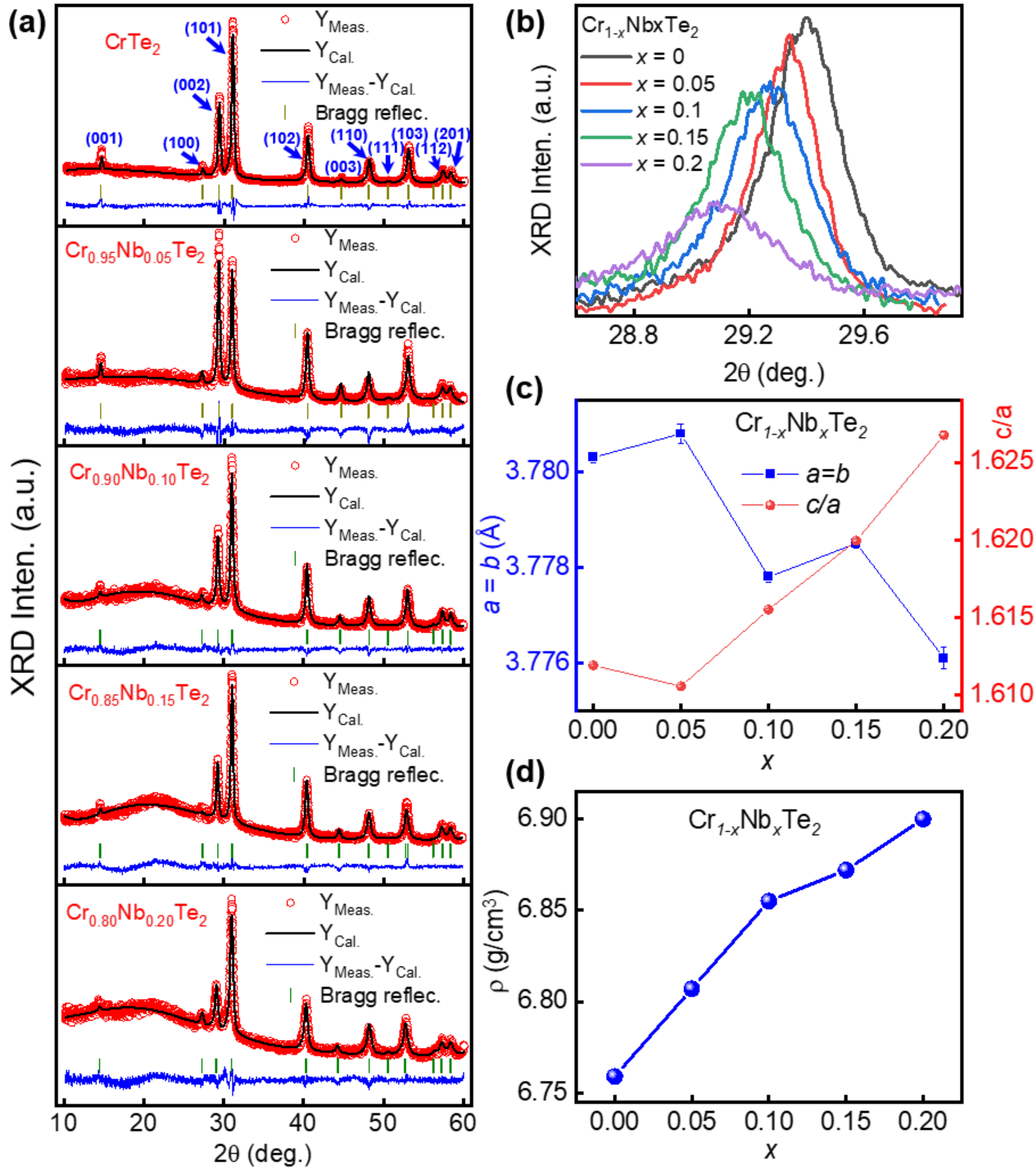

**Figure S1.2.1.** (a) Measured (scattered) and fitted (solid line) XRD spectroscopy for 1*T*-$Cr_{1-x}Nb_xTe_2$ ($x$ = 0, 0.05, 0.1, 0.15, and 0.2) crystals. The fitting was refined by using the Le Bail method with $P\bar{3}m1$ symmetry, summarized in Table S1. (b) The XRD peak at 2θ ~29°, corresponding to (002) plane, shifts towards lower diffraction angles with increasing Nb doping concentration. (c) The variation of the lattice parameters $a$ and $b$, and the ratio $c/a$, as a function of the Nb concentration. (d) Deduced density of the $Cr_{1-x}Nb_xTe_2$ crystals versus the Nb doping factor in the range $x = 0 - 0.2$.

Table S.2 contains the values of Miller indices (*h*, *k*, and *l*) of the different planes, XRD angle 2θ, and interplane spacing ($d_{hkl}$) values of $Cr_{0.8}Nb_{0.2}Te_2$ crystal. The red rows in Table S2 show the information about the planes mentioned in the main text.

**Table S2:** Deducing the (*hkl*) planes from the XRD peaks fitting

| No. | Code | *h* | *k* | *l* | Mult | $H_w$ | ETA/M | 2θ/TOF | $I_{calc.}$ | $I_{obs.}$ | Sigma | $StrFactor^2$ | $d_{hkl}$ | Corr. |
|---|---|---|---|---|---|---|---|---|---|---|---|---|---|---|
| 1 | 1 | 0 | 0 | 1 | 2 | 0.43232 | 0.4451 | 14.407 | 0 | 0 | 0 | 0 | 6.14281 | 1.000 |
| 2 | 1 | 1 | 0 | 0 | 6 | 0.47076 | 0.68609 | 27.247 | 39.5 | 39.9 | 2.573 | 424.1077 | 3.27024 | 1.000 |
| 3 | 1 | 0 | 0 | 2 | 2 | 0.47794 | 0.71989 | 29.049 | 159.9 | 160.5 | 1.94 | 5912.7222 | 3.07141 | 1.000 |
| 4 | 1 | 0 | 1 | 1 | 6 | 0.48599 | 0.75563 | 30.953 | 226.6 | 227.4 | 1.642 | 3201.5139 | 2.88666 | 1.000 |
| 5 | 1 | 1 | 0 | 1 | 6 | 0.48599 | 0.75563 | 30.953 | 226.6 | 227.4 | 1.642 | 3201.5139 | 2.88666 | 1.000 |
| 6 | 1 | 0 | 1 | 2 | 6 | 0.53204 | 0.9301 | 40.249 | 106 | 106.3 | 1.324 | 2664.3362 | 2.23881 | 1.000 |
| 7 | 1 | 1 | 0 | 2 | 6 | 0.53204 | 0.9301 | 40.249 | 106 | 106.3 | 1.324 | 2664.3362 | 2.23881 | 1.000 |
| 8 | 1 | 0 | 0 | 3 | 2 | 0.55494 | 1.00417 | 44.195 | 38.4 | 38.5 | 1.107 | 3575.6582 | 2.0476 | 1.000 |
| 9 | 1 | 1 | 1 | 0 | 6 | 0.57994 | 1.07848 | 48.155 | 163.5 | 164 | 1.762 | 6159.2036 | 1.88807 | 1.000 |
| 10 | 1 | 1 | 1 | 1 | 12 | 0.59592 | 1.12306 | 50.53 | 14.6 | 14.5 | 0.46 | 305.9398 | 1.80475 | 1.000 |
| 11 | 1 | 0 | 1 | 3 | 6 | 0.61116 | 1.16376 | 52.699 | 98.2 | 98.7 | 0.954 | 4540.4829 | 1.73548 | 1.000 |
| 12 | 1 | 1 | 0 | 3 | 6 | 0.61116 | 1.16376 | 52.699 | 98.2 | 98.7 | 0.954 | 4540.4829 | 1.73548 | 1.000 |
| 13 | 1 | 2 | 0 | 0 | 6 | 0.6372 | 1.22965 | 56.209 | 21.9 | 21.9 | 0.454 | 1183.1703 | 1.63512 | 1.000 |
| 14 | 1 | 1 | 1 | 2 | 12 | 0.64505 | 1.24873 | 57.226 | 80.9 | 81.5 | 0.996 | 2301.7917 | 1.60847 | 1.000 |
| 15 | 1 | 0 | 2 | 1 | 6 | 0.65393 | 1.26985 | 58.351 | 42 | 42 | 0.497 | 2577.6838 | 1.5801 | 1.000 |
| 16 | 1 | 2 | 0 | 1 | 6 | 0.65393 | 1.26985 | 58.351 | 42 | 42 | 0.497 | 2577.6838 | 1.5801 | 1.000 |

We also performed single-crystal X-ray diffraction (SCXRD) on 1*T*-$CrTe_2$, $Cr_{0.9}Nb_{0.1}Te_2$, and $Cr_{0.8}Nb_{0.2}Te_2$ compositions, as shown in Figure S.1.2.2. A clear shift of the XRD peaks towards lower angles (see the inset of Figure S.1.2.2 at 2θ ~30.9°), confirming the systematic expansion of the lattice parameter *c* upon Nb doping. *c* is 6.0252 Å for 1*T*-$CrTe_2$, 6.1226 Å for $Cr_{0.9}Nb_{0.1}Te_2$, and 6.1466 Å for $Cr_{0.8}Nb_{0.2}Te_2$, respectively. The SCXRD agrees well with powder XRD (PXRD) results.

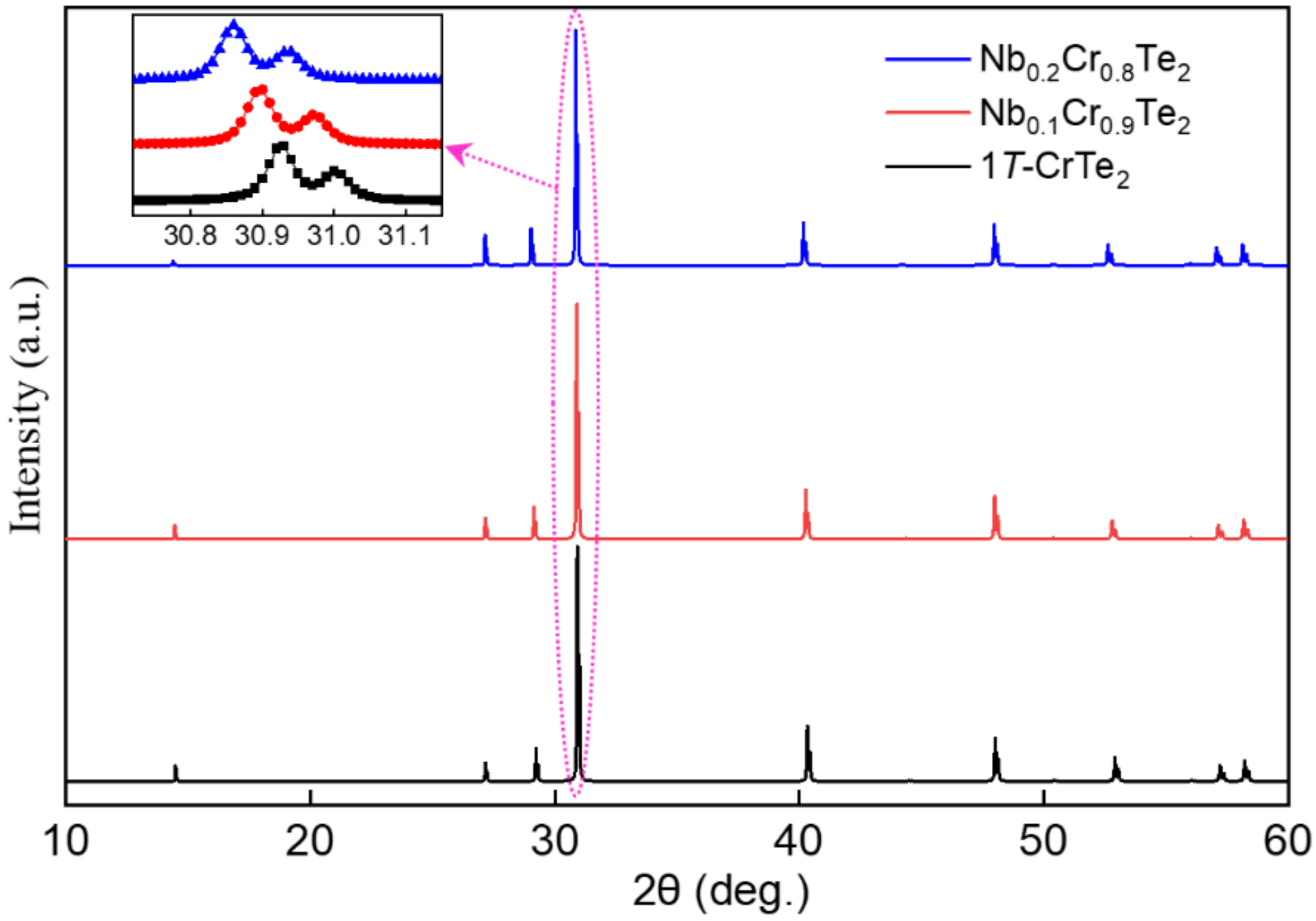

**Figure S1.2.2.** Measured SCXRD spectroscopy of 1*T*-$CrTe_2$, $Cr_{0.9}Nb_{0.1}Te_2$, and $Cr_{0.8}Nb_{0.2}Te_2$ compositions. A clear shift of the XRD peaks towards lower angles as shown in the inset at 2θ ~30.9°.

PXRD measurements on as-synthesized fresh $Cr_{0.90}Nb_{0.10}Te_2$ and on $CrTe_2$ crystals after exposure to ambient conditions for 24 hrs show a good stability (see Figure S1.2.3).

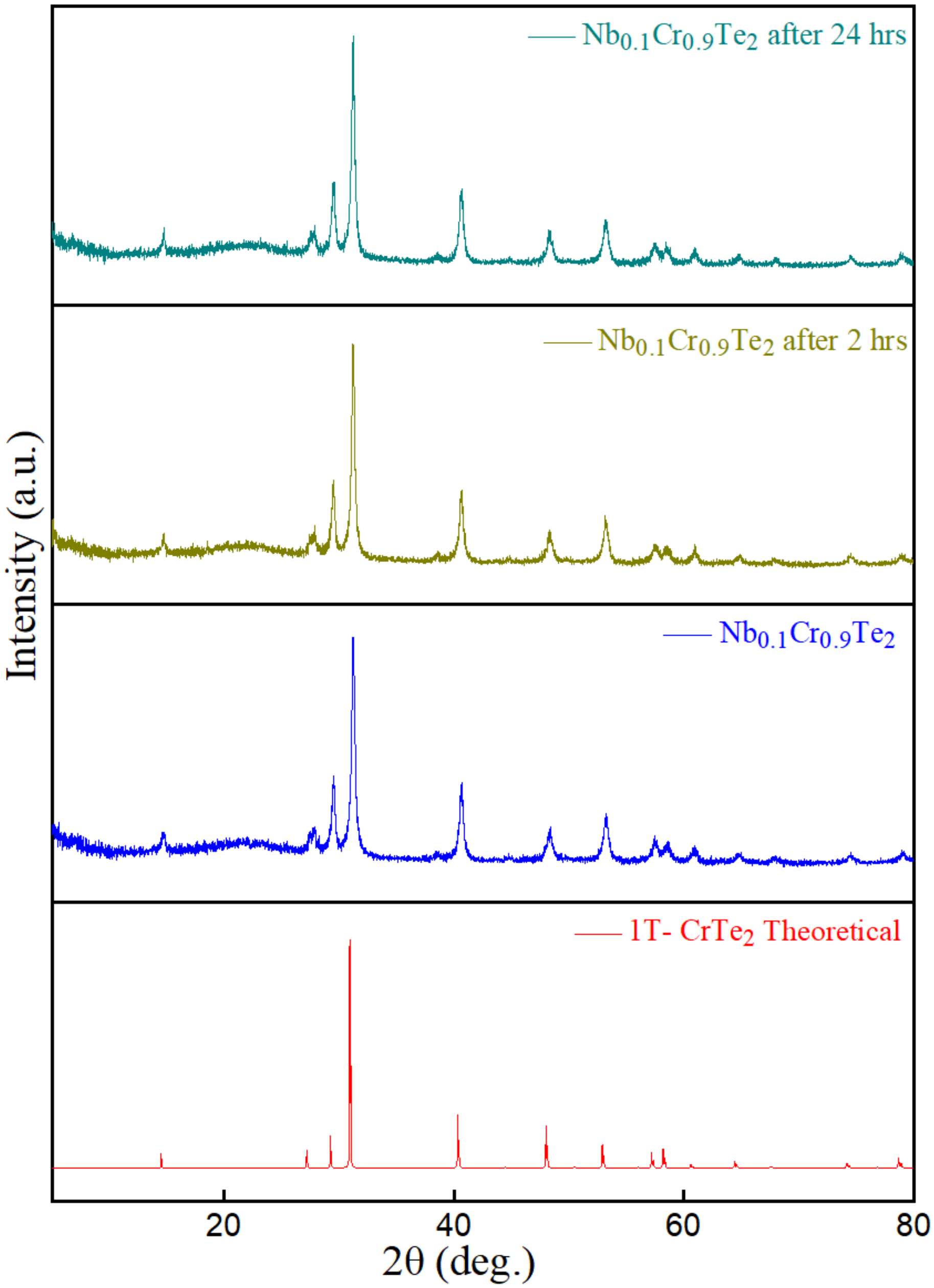

**Figure S1.2.3.** PXRD spectra of as-synthesized $Cr_{0.90}Nb_{0.10}Te_2$ and on $CrTe_{2\ crystals}$ after exposure to ambient conditions for 24 hrs

### S1.3 Transmission electron microscopy of 1*T*-$Cr_{0.8}Nb_{0.2}Te_2$ flakes

We performed high resolution transmission electron microscopy (HRTEM) on selected 1*T*-$Cr_{0.8}Nb_{0.2}Te_2$ flakes, see Figure S1.3. Variations in lattice orientation, contrast, and fringe continuity indicate a polycrystalline structure composed of numerous nanoscale domains rather than a single uniform crystal.

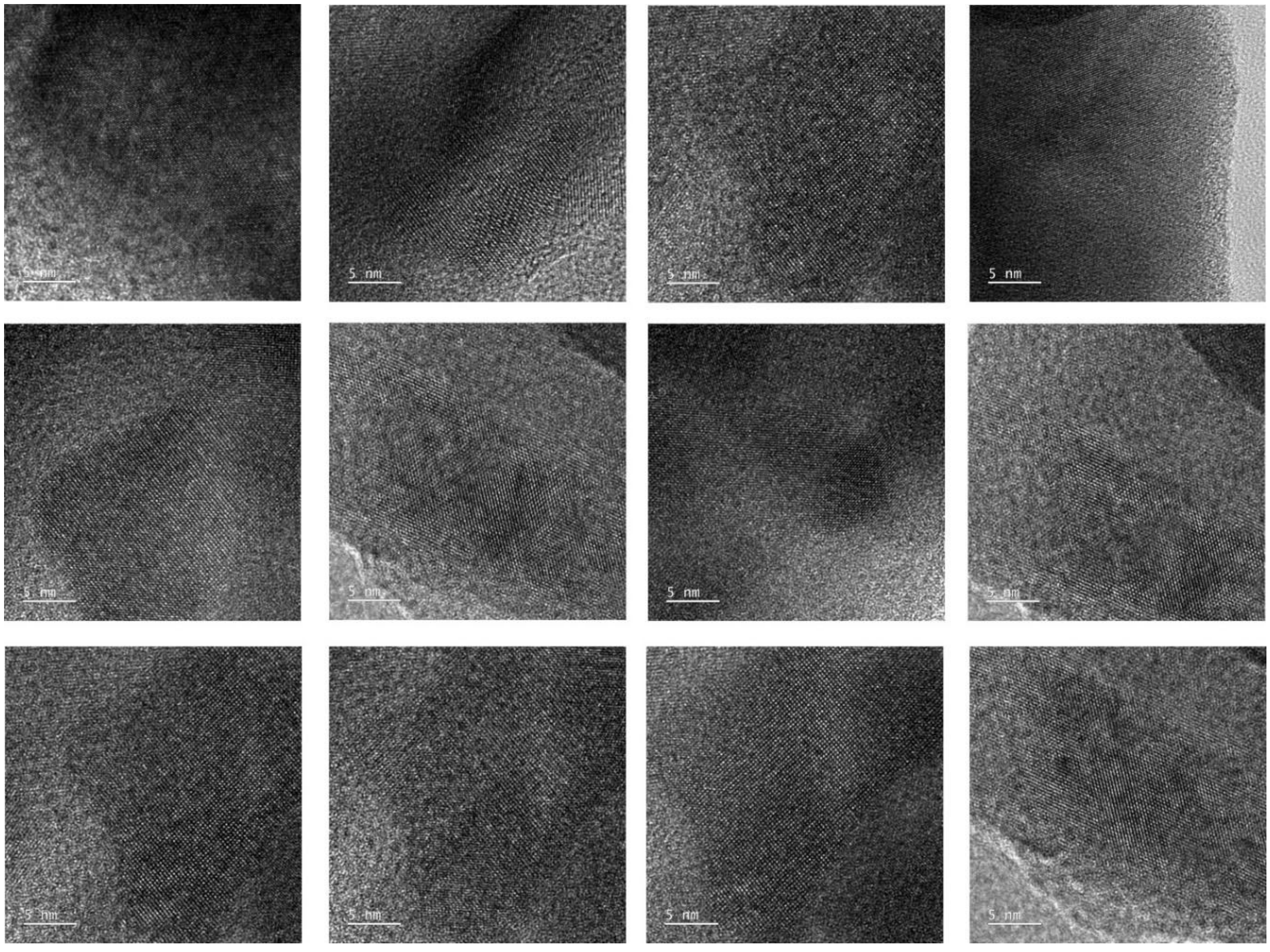

**Figure S1.3.1.** HRTEM images of 1*T*-$Cr_{0.8}Nb_{0.2}Te_2$ nanoflakes.

Zoomed HRTEM image of one of the $Cr_{0.8}Nb_{0.2}Te_2$ nanoflakes in Figure S1.3.2 confirms the well-resolved (100) lattice fringes consistent with the expected interplanar spacing (d ≈0.32 nm).

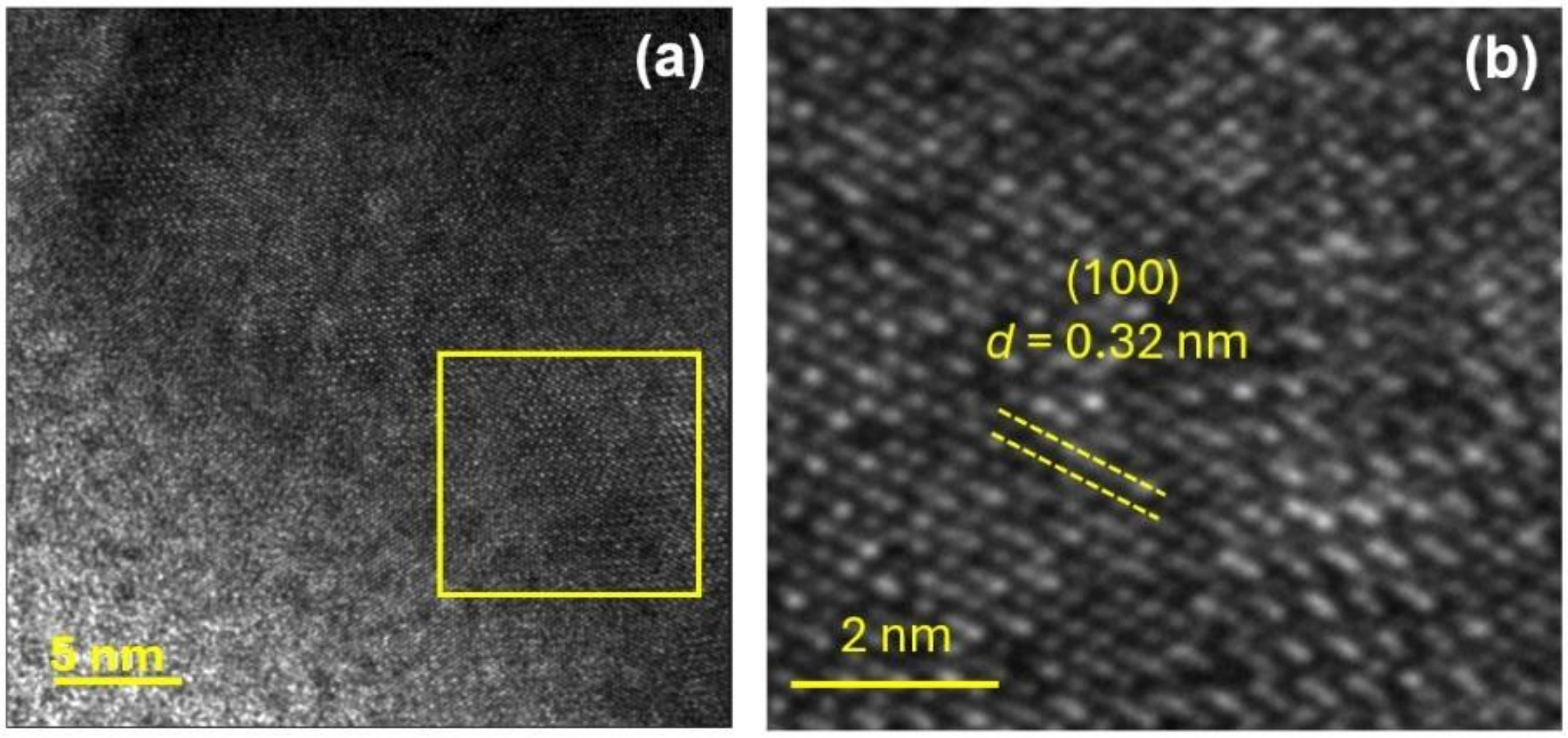

**Figure S1.3.2.** (a) HRTEM image of 1*T*-$Cr_{0.8}Nb_{0.2}Te_2$ crystal with different crystal orientations. (b) Zoomed HRTEM image of the region in (a) highlighted by a solid square showing the (100) lattice fringes.

## S1.4 X-ray photoelectron spectroscopy of $Cr_{1-x}Nb_xTe_2$ crystals

We measured X-ray photoelectron spectroscopy (XPS) spectra on selected $Cr_{1-x}Nb_xTe_2$ crystals utilizing Thermo Scientific Al K-Alpha XPS at a pressure of $\sim 10^{-9}$ mbar.[45,57] Further details about XPS measurements can be found in our previous works.[45,57] All spectra were referenced to C1s peak at an energy of 284.8 eV to remove any residual charging. Figure S1.4(a-b) shows XPS peaks of Te and Cr in pristine and $Cr_{0.8}Nb_{0.2}Te_2$ crystals, respectively. The binding energies of Te ($3d_{3/2}$, $3d_{5/2}$) and Cr ($2p_{1/2}$, $2p_{3/2}$) doublets are (582.8 eV, 572.5 eV), and (586.4 eV, 575.8 eV), respectively, consistent with the findings in Reference [35].

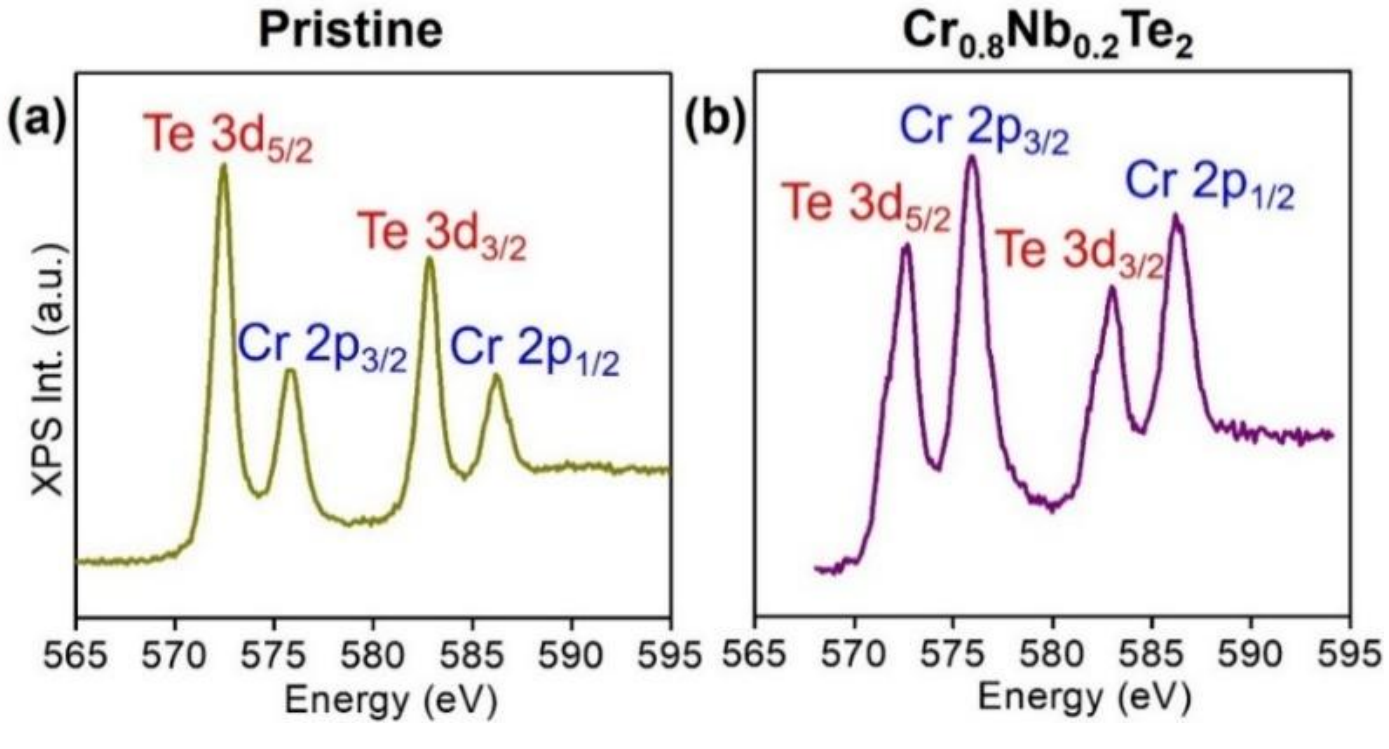

**Figure S1.4.** XPS spectra of 1*T*-$CrTe_2$ (a) and 1*T*-$Cr_{0.8}Nb_{0.2}Te_2$ (b).

## S1.5 Energy-dispersive X-ray spectroscopy measurements

To verify the niobium incorporation into the $CrTe_2$ lattice, we performed energy-dispersive X-ray (EDS) mapping by using scanning transmission electron microscopy (STEM) and high-angle annular dark-field (HAADF) imaging on selected thin $Cr_{0.8}Nb_{0.2}Te_2$ nanoflakes. The resulting elemental maps confirm the successful mixing of Nb within the $CrTe_2$ crystal, see Figure S1.5.

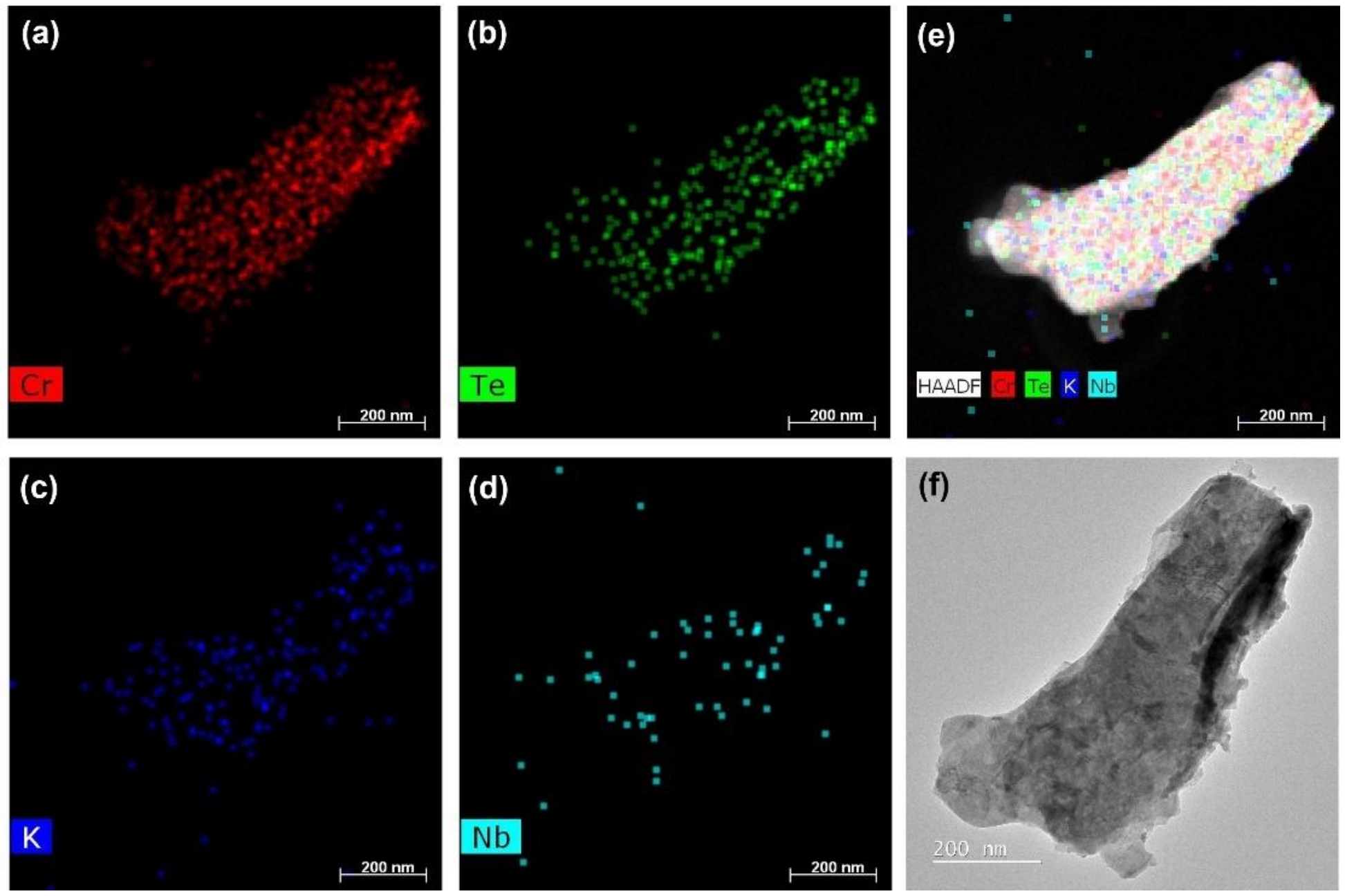

**Figure S1.5.** (a–d) EDS elemental maps of Cr, Te, K, and Nb in 1*T*-$Cr_{0.8}Nb_{0.2}Te_2$ nanocrystal, each displayed in a separate color channel to highlight their spatial distribution. (e) Overlapped Cr, Te, K, and Nb HAADF-STEM/EDS maps confirm a uniform elemental dispersion within the $CrTe_2$ crystal. (f) HAADF-STEM image of the analyzed $Cr_{0.8}Nb_{0.2}Te_2$ nanoflake.

### S1.6 Electron energy loss spectroscopy

To check the distribution of niobium atoms in the $CrTe_2$ lattice, we performed additional structural imaging scanning using transmission electron microscopy (STEM, Titan ETEM 300 keV) and electron energy loss spectroscopy (EELS) on selected 1*T*-$Cr_{0.8}Nb_{0.2}Te_2$ nanoflakes. was performed using a Titan ETEM operating at 300 keV. The EELS measurements were made using a Gatan Biocontinuum/K3-IS. Figure S1.6 confirms a homogeneous distribution of Cr, Te, and Nb within the crystal. Due to the overlap of Cr and Te, it is hard to extract the exact elemental concentrations, and additional deconvolution and principal component analysis are needed.[80]

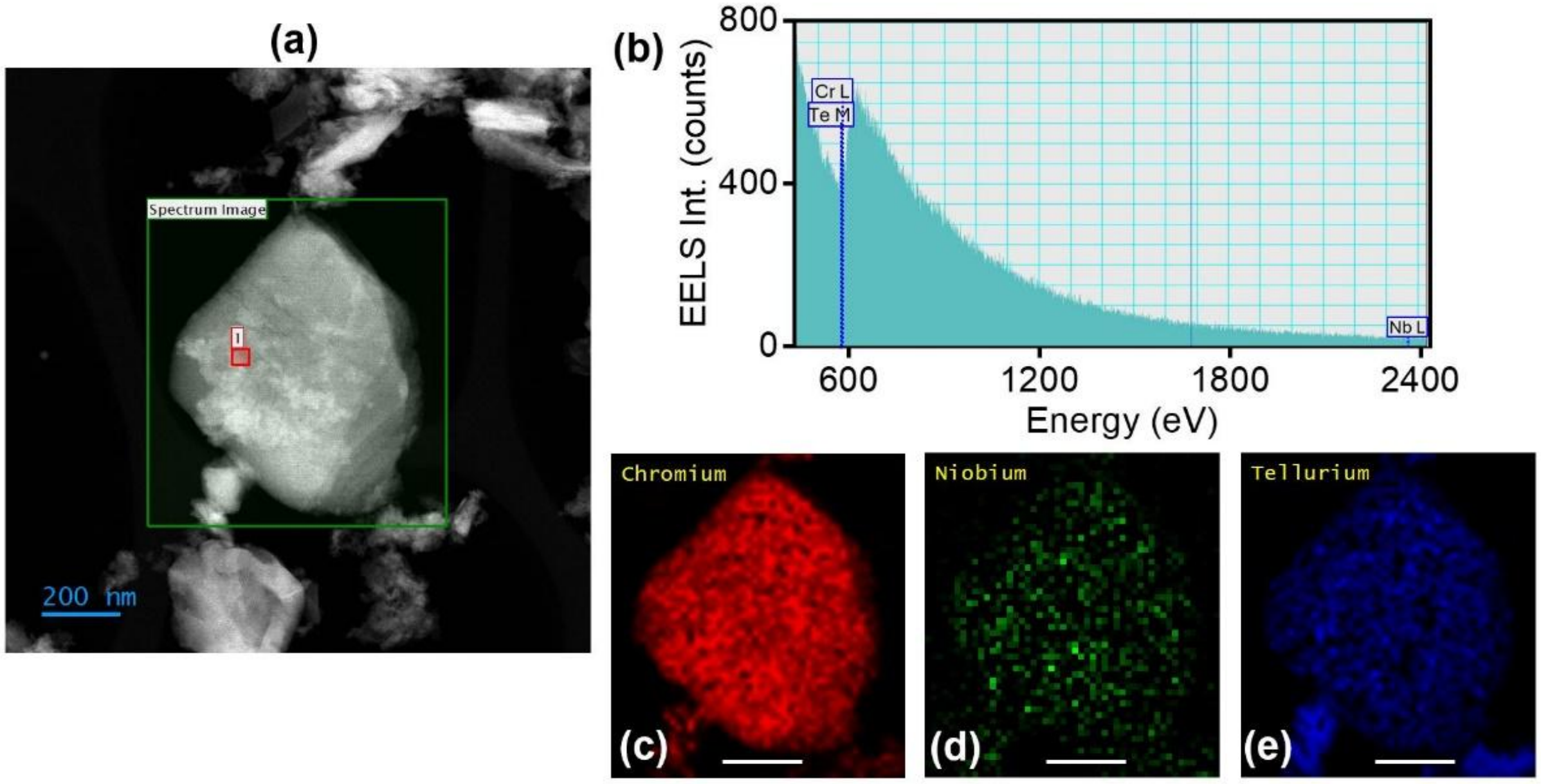

**Figure S1.6. (a)** STEM-HAADF image of 1*T*-$Cr_{0.8}Nb_{0.2}Te_2$ nanoflake. (b) EELS spectrum showing the overlap of Cr and Te elements. (c-e) Element maps of Cr, Nb, and Te, respectively.

### Section S2: Additional magnetic measurements on $Cr_{1-x}Nb_xTe_2$ ($x = 0 – 0.2$) crystals

As discussed in the main text, we performed multiple runs of magnetic measurements on the 1*T*-$Cr_{1-x}Nb_xTe_2$ ($x = 0 – 0.2$) crystals to deduce the average values of the Curie temperature ($T_C$) and saturation magnetizations (*Ms*). All the compositions exhibit ferromagnetic behavior with $T_C$ values in the range of 315 – 295 K (see Figures S2.1, S2.2, S2.3, S2.4, and S2.5). The parent $CrTe_2$ and lower Nb-doped concentrations up to 15% show a good stability of the magnetic properties even after 387 days. The variation in $T_C$ and $M_S$ for $Cr_{0.9}Nb_{0.1}Te_2$ crystals is explained by the averaged measurements with different crystal sizes that lead to anisotropic behavior. However, a variation in $T_C$ and $M_S$ is apparent in some compositions, this can be explained by the instability of the 20% Nb doped crystals, size of the microcrystals (microcrystals vs nanoflakes), or control of the Nb composition of the initial 10% Nb grown crystals. We note that the $Cr_{0.9}Nb_{0.1}Te_2$ and $Cr_{0.8}Nb_{0.2}Te_2$ powders (nanoflakes) were obtained by crashing the big freshly made crystals. While there are no structural changes observed from EELS and EDS (discussed above) in the $Cr_{0.8}Nb_{0.2}Te_2$ powder, its magnetic properties are slightly affected, manifested by a small change in $T_C$ for $Cr_{0.9}Nb_{0.1}Te_2$ (~316 K for crystal vs ~313.5 K for powder) and $Cr_{0.8}Nb_{0.2}Te_2$ (~300.67 K for crystal vs ~300.37 K for powder). However, we see an increase of the coercive field $H_C$ (insets of Figures S2.3c and S2.5c) in $Cr_{0.9}Nb_{0.1}Te_2$ (~1 mT of crystal to ~5 mT for powder) and in $Cr_{0.8}Nb_{0.2}Te_2$ (~1 mT of crystal to ~9 mT for powder), that is may be explained by size effects.[81,82]

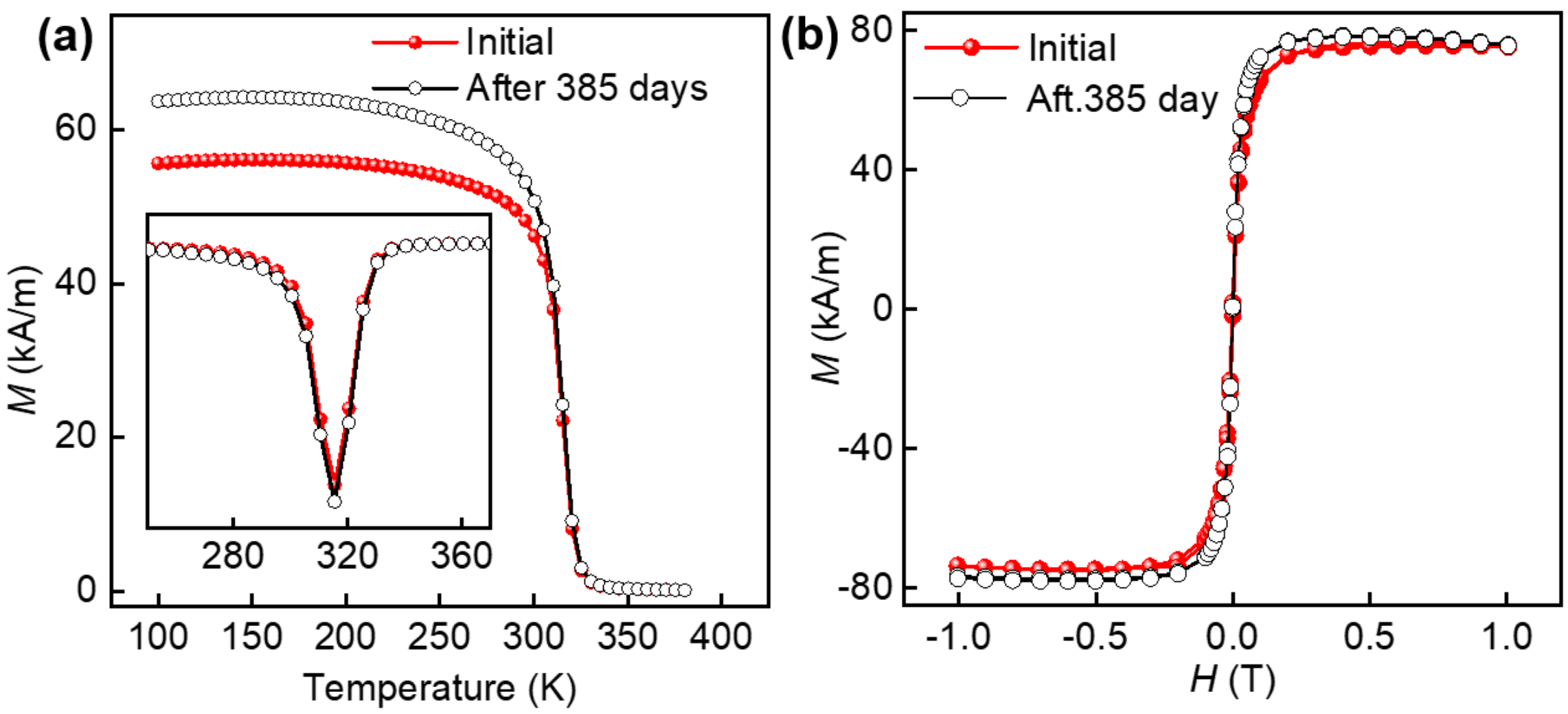

**Figure S2.1.** Magnetometry of 1*T*-$CrTe_2$ crystals. (a) In-plane magnetization (*M*) vs temperature (*T*) curves at an external magnetic field of 30 mT. Inset of (a): Derivative of M-H curves in (a) with $T_C$ of 314.88 ± 0.15 K and 314.75 ± 0.28 K. The average $T_C$ is 314.79 K. (b) In-plane *M-H* hysteresis loops measured at 300 K. The average $M_S$ is 77.19 kA/m (76.68 kA/m for the first measurements and 77.69 kA/m for the repeated measurements after 385 days.

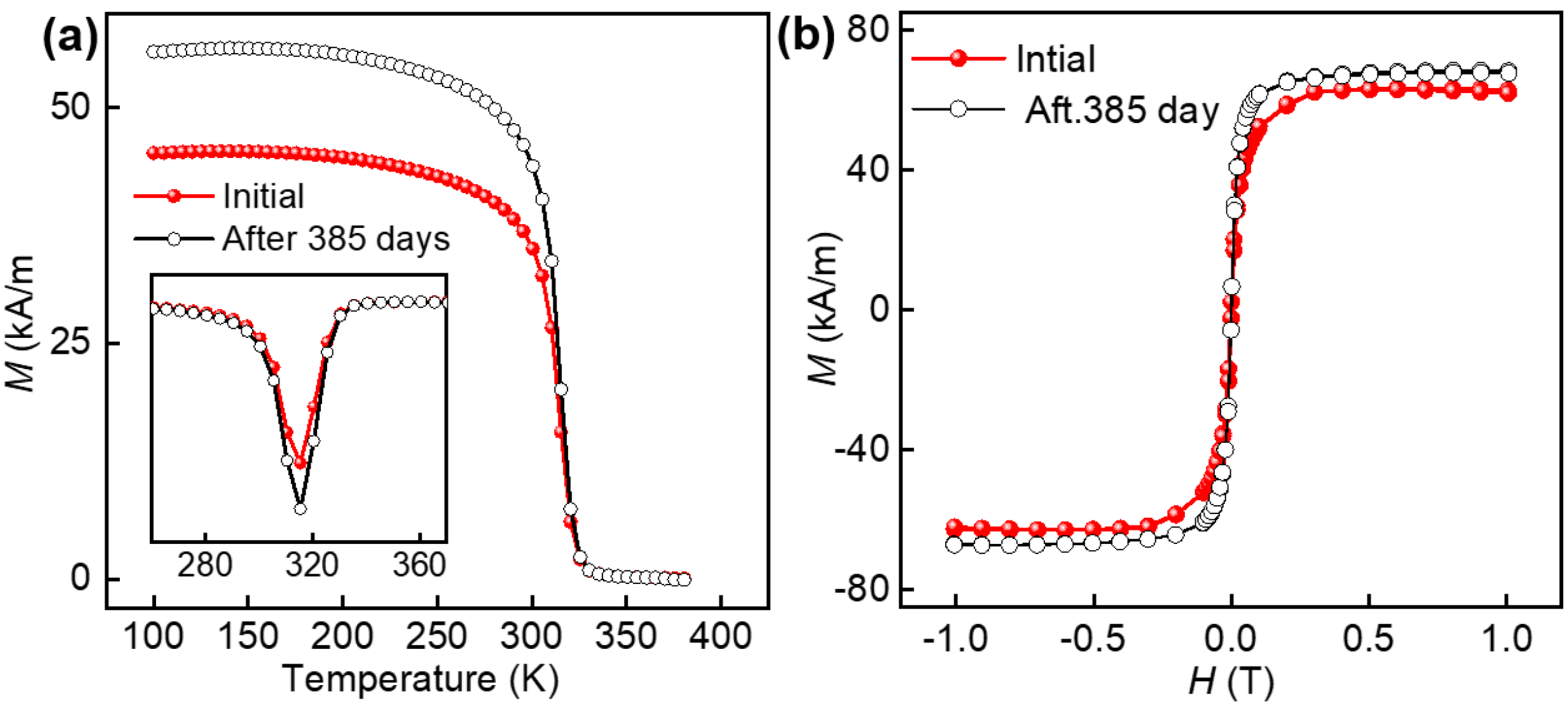

**Figure S2.2.** Magnetometry of 1*T*-$Cr_{0.95}Nb_{0.05}Te_2$ crystals. (a) In-plane magnetization (*M*) vs temperature (*T*) curves at an external magnetic field of 30 mT. Inset of (a): Derivative of M-H curves in (a) with $T_C$ of 314.18 ± 0.28 K and 314.49 ± 0.3 K. The average $T_C$ is 314.34 K. (b) In-plane *M-H* hysteresis loops measured at 300 K. The average $M_S$ is 65.66 kA/m (62.86 kA/m for the first measurements and 68.46 kA/m repeated measurements after 385 days.

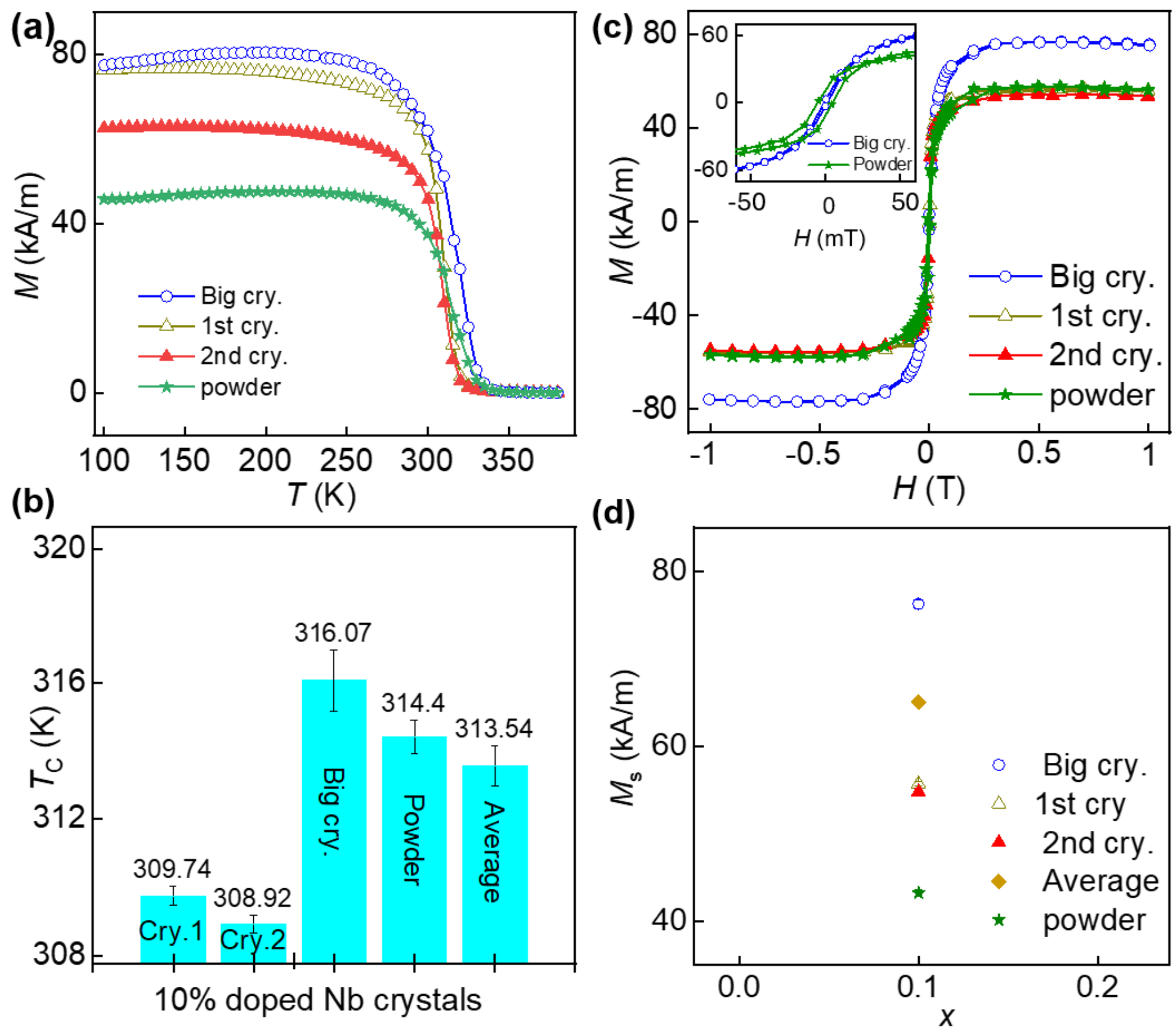

**Figure S2.3.** Magnetometry of 1*T*-$Cr_{0.9}Nb_{0.1}Te_2$ crystals. (a) In-plane magnetization (*M*) vs temperature (*T*) curves at an external magnetic field of 30 mT. (b) In-plane *M*-*H* hysteresis loops measured at 300 K. $T_C$ (c) and saturation magnetization $M_S$ (d) for measurements on powder (nanoflakes) and bulk crystals with different batches. The average $T_C$ is 314.34 K and the average $M_S$ is 65.01 kA/m.

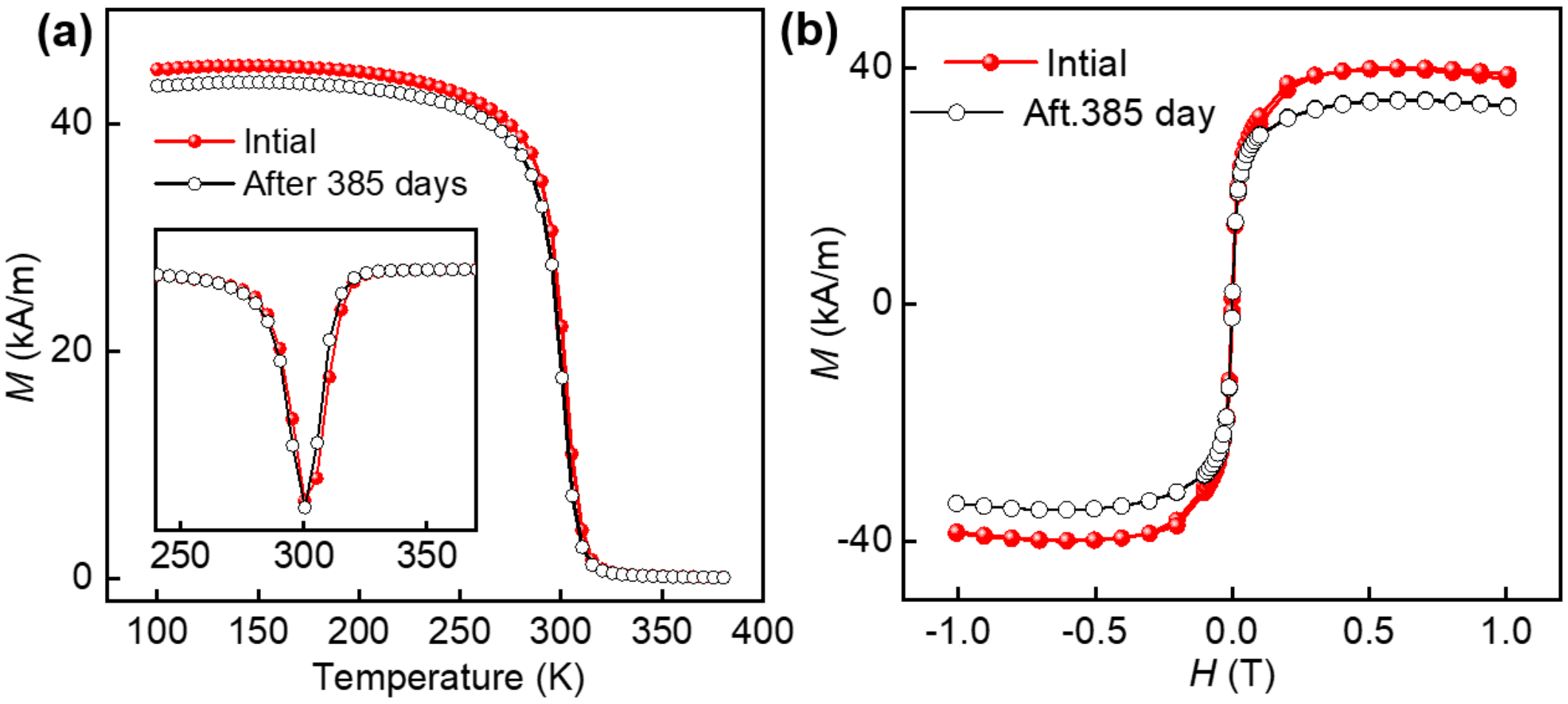

**Figure S2.4.** Magnetometry of 1*T*-$Cr_{0.85}Nb_{0.15}Te_2$ crystals. (a) In-plane magnetization (*M*) vs temperature (*T*) curves at an external magnetic field of 30 mT. Inset of (a): Derivative of M-H curves in (a) with $T_C$ of 301.54 ± 0.2 K and 299.96 ± 0.25 K. The average $T_C$ is 300.75 K. (b) In-plane *M*-*H* hysteresis loops measured at 300 K. The average $M_S$ is 37.34 kA/m (40.02 kA/m for the first measurements and 34.66 kA/m repeated measurements after 385 days.

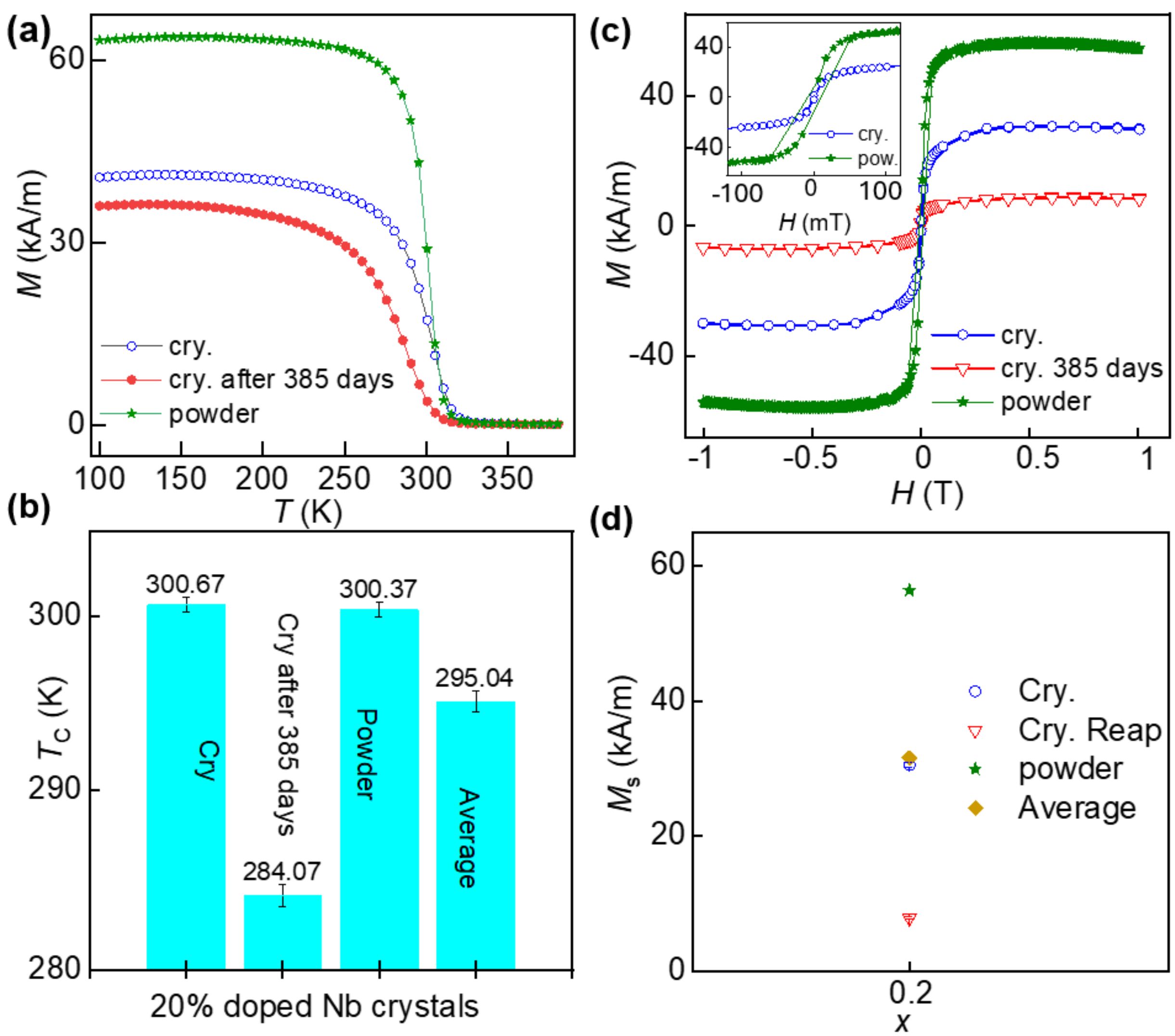

**Figure S2.5.** Magnetometry of 1$T$-$Cr_{0.8}Nb_{0.2}Te_2$ crystals. (a) In-plane magnetization ($M$) vs temperature ($T$) curves at an external magnetic field of 30 mT. (b) In-plane $M$-$H$ hysteresis loops measured at 300 K. $T_C$ (c) and saturation magnetization $M_S$ (d) for measurements on powder (nanoflakes) and bulk crystals with different batches. The average $T_C$ is 295.04 K and the average $M_S$ is 31.58 kA/m.

Figure S.2.6 depicts $T_C$, the magnetic anisotropy ($H_A$) extracted from Ferromagnetic resonance spectroscopy (FMR) spectroscopy, and the $c/a$ ratio deduced from XRD refinements for $Cr_{1-x}Nb_xTe_2$ ($x = 0.05 - 0.2$) crystals. $T_C$, $H_A$, and $c/a$ change slightly until $x = 0.1$. Then, they change drastically for Nb concentration $x \geq 10\% - 20\%$, that is closely related to the effect of the increasing c/a ratio upon replacing the small (0.55 Å) Cr atoms with larger (0.68 Å) Nb atoms.[32] Generally, in layered materials, in-plane bonds are much stiffer and stronger compared to the weak van der Waals forces that bind the layers together. The lattice strain from the larger Nb atoms is thus more easily accommodated by an expansion of the interlayer spacing (the c-axis parameter). The expansion of c/a ratio also alters the environment around the Te and Cr atoms, which changes the spin-orbit coupling.[35] Since spin-orbit coupling strongly influences the magnetic anisotropy amplitude, it is a key parameter that also affects the Curie temperature (discussed below in Section S3).

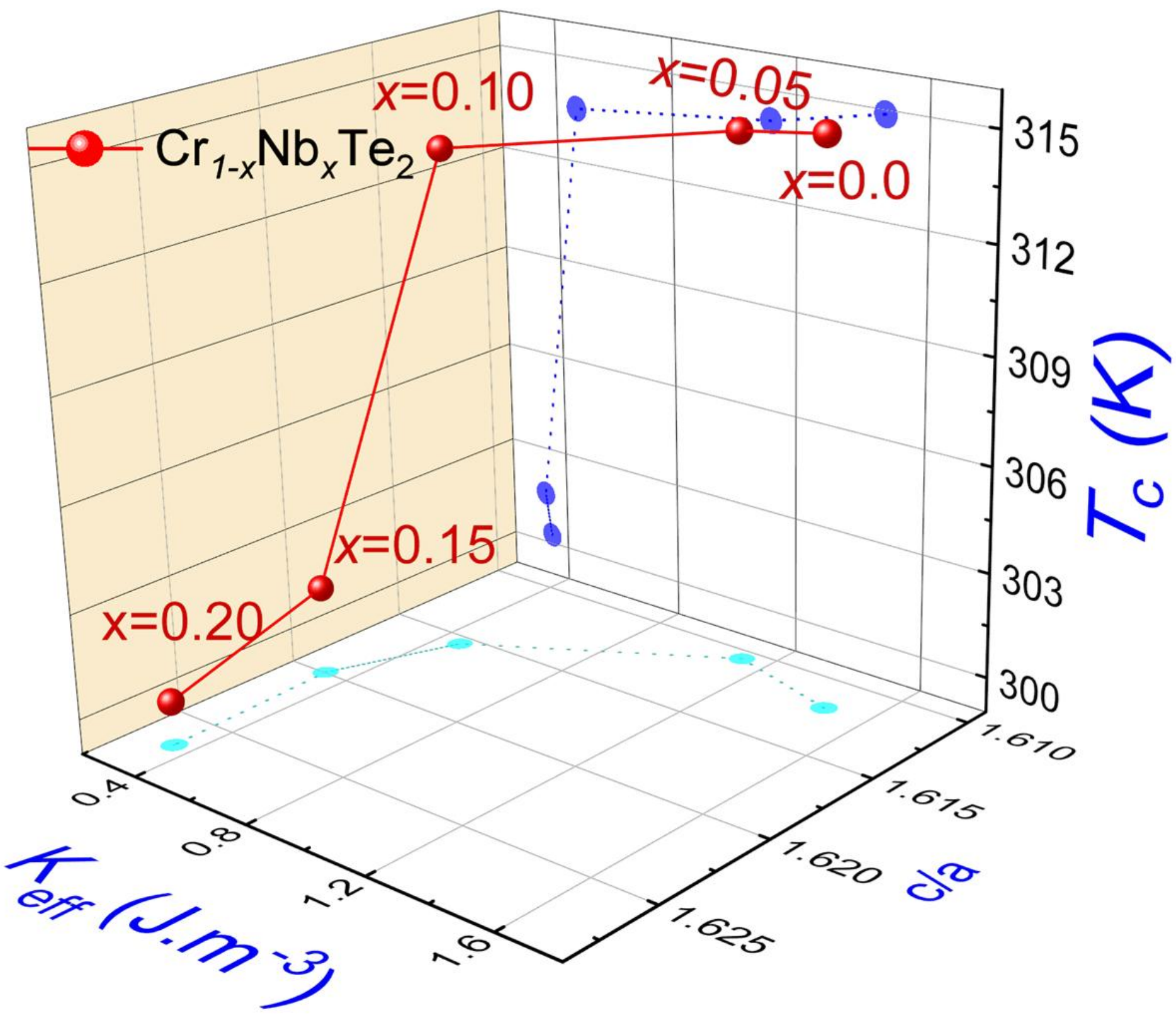

**Figure S2.6.** 3D representation of $T_C$, $K_{eff}$, c/a ratio, and Nb doping parameter $x$.

Figure S2.7 shows in-plane (IP) and out-of-plane (OPP) bulk magnetic measurements: M-H and FMR intensity vs H on pristine 1*T*-$CrTe_2$ crystals. which confirms in-plane magnetic anisotropy, as discussed in the main text.

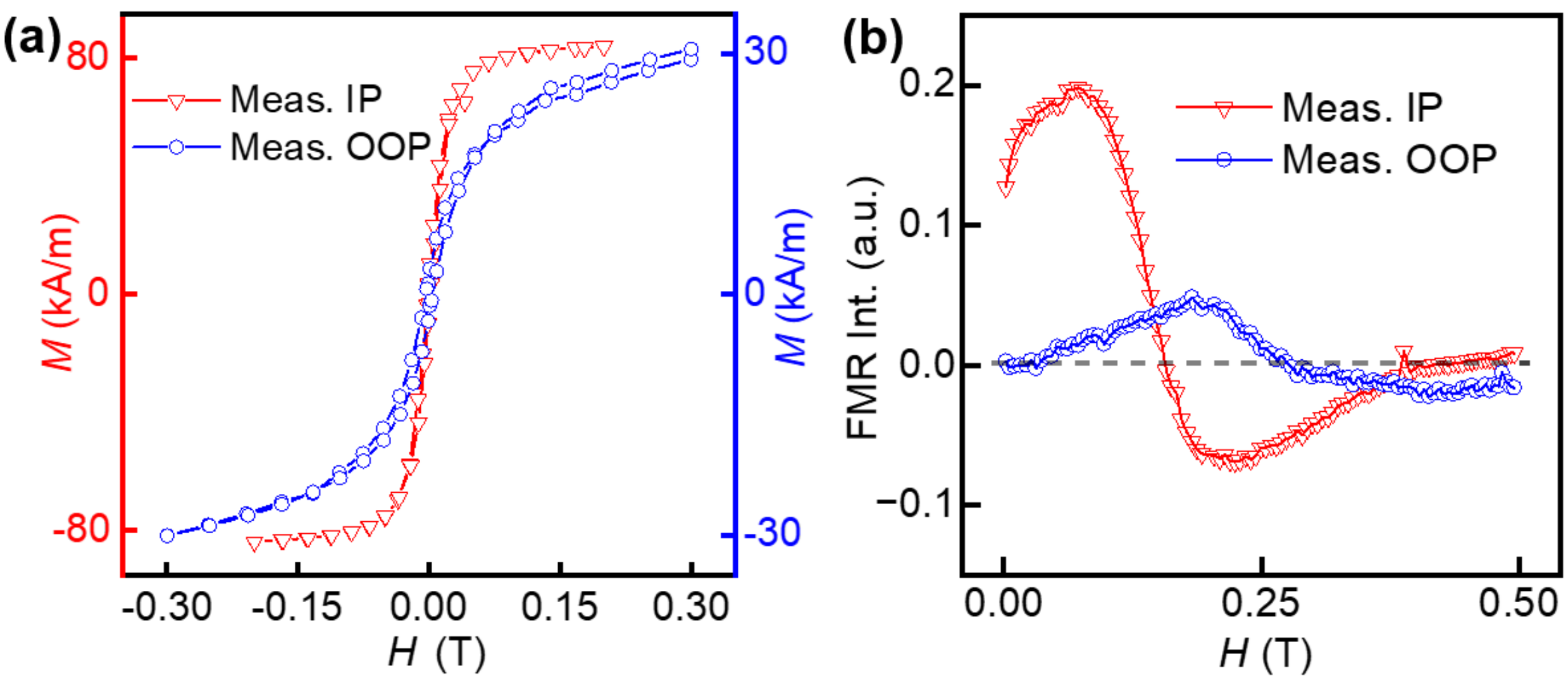

**Figure S2.7.** (a) IP (triangles) and OOP (circles) magnetization ($M$) versus applied magnetic field ($H$) curves at a temperature of 300 K of pristine $CrTe_2$ crystals. (b) IP (triangles) and OOP (circles) FMR intensity vs $H$ at a MW frequency of 17.5 GHz at a temperature of 297 K of pristine 1*T*-$CrTe_2$ crystals.

## Section S3: Ferromagnetic resonance measurements on pristine 1*T*-$CrTe_2$ flakes

FMR spectroscopy was performed on 1*T*-$CrTe_2$ flakes in a helium-filled cryostat to prevent oxidation. We used flip chip FMR method to deduce the frequency of magnetic resonance, magnetic anisotropy, and damping constant $\alpha$.[45,66] First, the flakes with a varying thickness were exfoliated onto double-side polished sapphire substrates using scotch tape exfoliation method.[67,68] The sapphire substrate with the 1*T*-$CrTe_2$ exfoliated flakes (facing down) was carefully placed over the microwave (MW) stripline to study their spin dynamics. Figure S3.1a shows the flakes on top of MW line. Due to the weak signal from individual flakes (not shown here), we measured the FMR response (Figure 5, main text) from multiple flakes on top of the MW line. Figures S3.1b(d) and S3.1c(d) show the optical images and corresponding atomic force microscope (AFM) height profiles of selected $CrTe_2$ flakes, highlighted by dashed arrows in Figure S3.1a. The flakes have a varying thickness in the range of 60 – 80 nm.

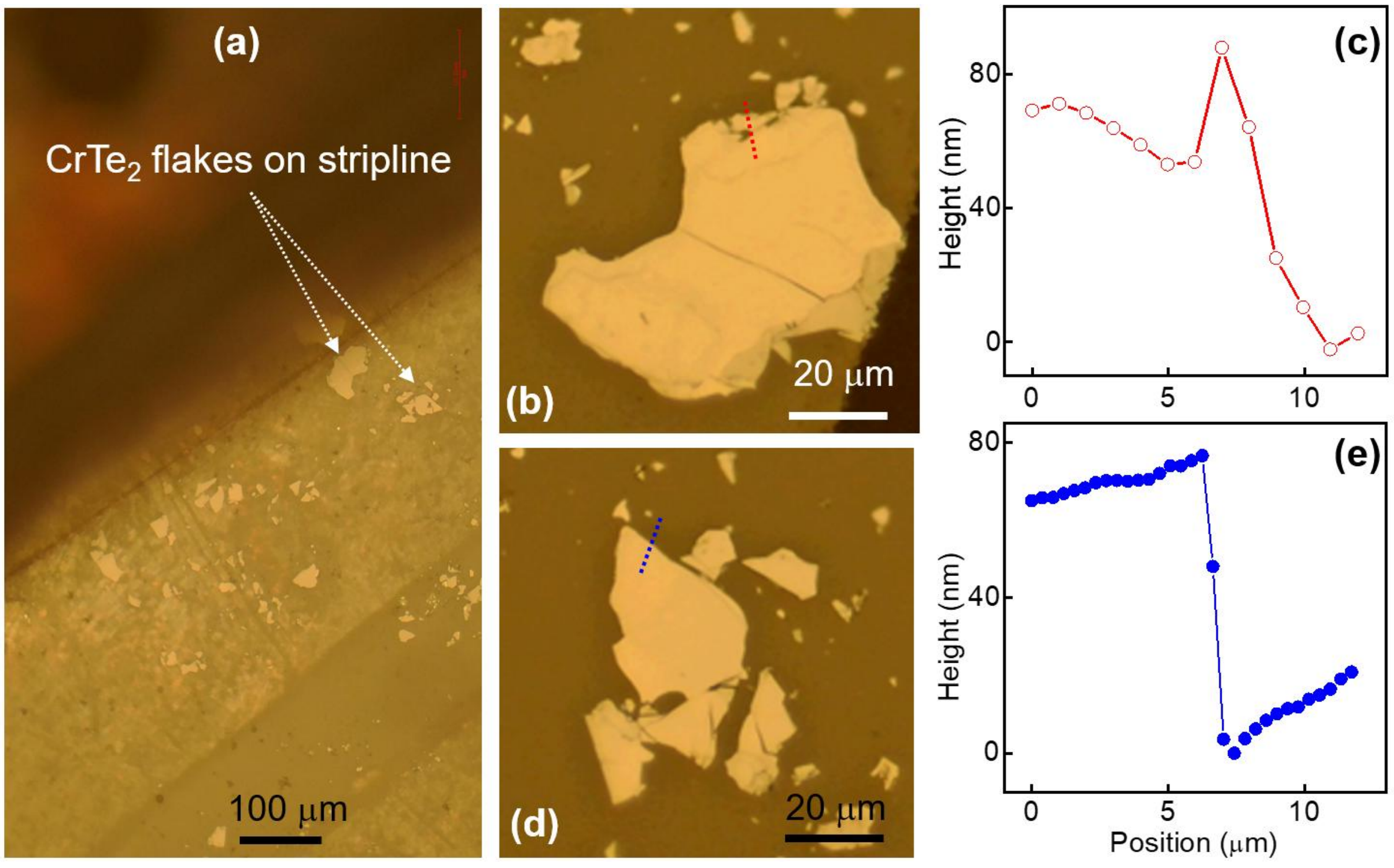

**Figure S3.1.** (a) Optical microscopy image of exfoliated 1*T*-$CrTe_2$ flakes on sapphire substrate, placed on top of MW stripline used to excite FMR. Optical images (b, d) and height profiles (c, e) of selected flakes obtained from AFM measurements.

Figure S3.2 shows the FMR response of the bulk 1*T*-$CrTe_2$ crystals and thin flakes exfoliated into sapphire (Figure S3.1a). The resonance field ($H_R$) is higher in bulk crystals due to the higher effective magnetic anisotropy ($K_{eff}$) of ~$1.47 \times 10^5$ J/m$^3$, compared to the value of ~ $0.4 \times 10^5$ J/m obtained on the flakes. While there is an increase of FMR linewidth $\Delta H$ for flakes (FigureS3.2b), the magnetic damping $\alpha$ is slightly lower (~ 0.0417) than the bulk crystals (~0.0665). Further discussion is detailed in the main text.

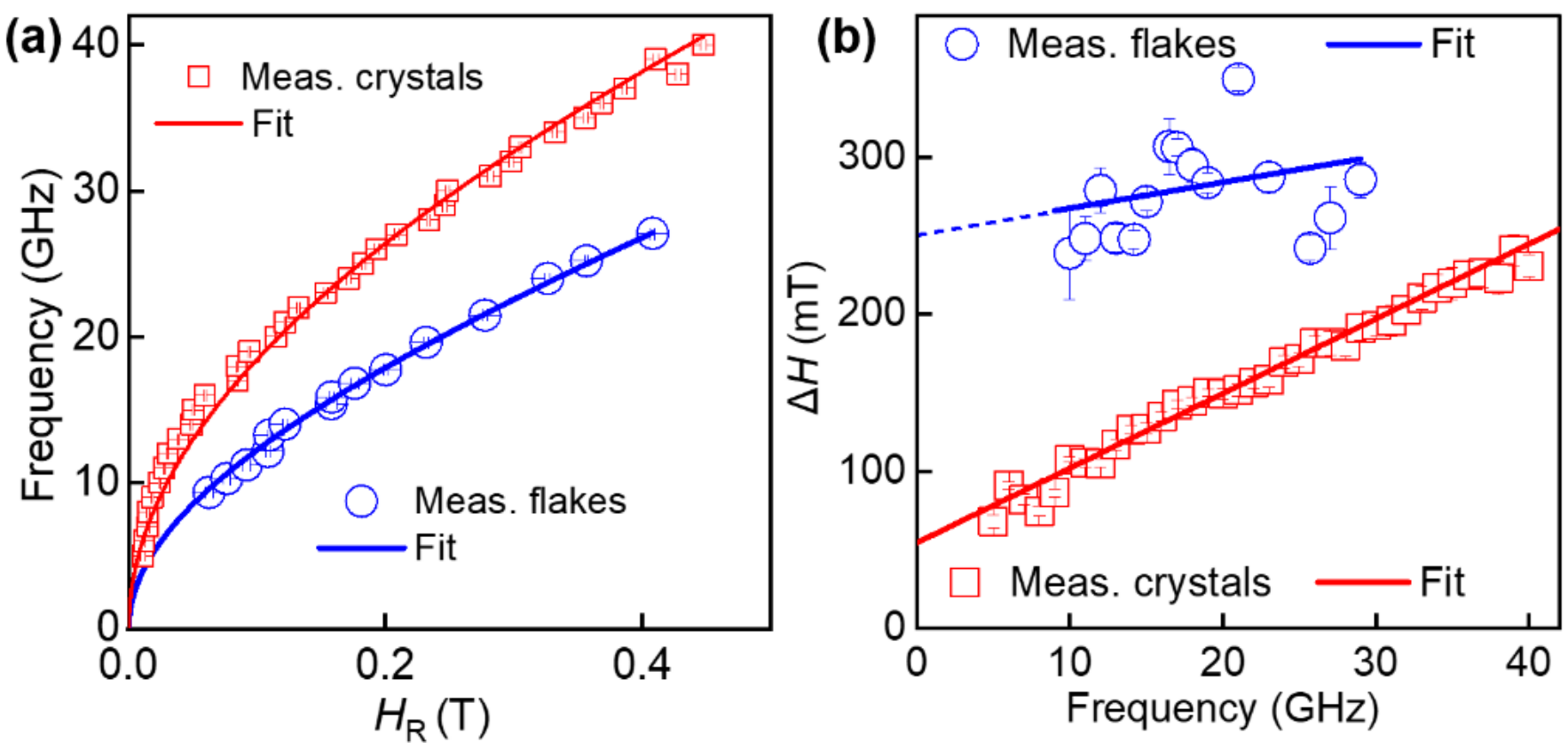

**Figure S3.2.** Comparison of magnetization dynamics of pristine 1*T*-$CrTe_2$ crystals and flakes. (a) FMR frequency as a function of field plot of bulk pristine 1*T*-$CrTe_2$ crystal and thin flakes (b) Extracted linewidths plotted vs frequency for bulk 1*T*-$CrTe_2$ crystal and thin flakes.

## Section S4: First-Principles Calculations and Spin Dynamics Simulations

The magnetic anisotropy energy of 1*T*-$Cr_{1-x}Nb_xTe_2$ is obtained from the internal energy difference between the OOP ($E_{out}$, Figure S4a) and IP ($E_{in}$, Figure S4b) spin configurations in the unit cell. The effective anisotropy coefficient $K_{eff}$ is deduced as:

$$K_{\text{eff}} = \frac{E_{\text{out}} - E_{\text{in}}}{V_{\text{uc}}},$$

where $V_{uc}$ is the unit cell volume.

To estimate the magnetic anisotropy at finite temperature (~297 K), we employed the Callen–Callen theory.[75] According to this theory, magnetic anisotropy scales with temperature as:

$$\frac{K(T)}{K(0)} \approx \left(1 - \frac{T}{T_c}\right)^{3\beta},$$

where $\beta$ is the critical exponent, which is typical ~0.33 for bulk 3D magnets. Using this value of $\beta$ and using $T_C$ taken from magnetic measurements, we estimated the magnetic anisotropy constant *K*. The calculated results are summarized in Table S4.1 and Figure 6f. As the temperature approaches $T_C$, the effective magnetic anisotropy is reduced by nearly one order of magnitude.

**Table S4.1.** Magnetic anisotropy of $Cr_{1-x}Nb_xTe_2$ at 0 K and room temperature

| Nb doping factor *x* | 0 | 0.05 | 0.1 | 0.15 | 0.2 |
|---|---|---|---|---|---|
| $T_c^{\text{Meas.}}$ (K) | **314.79** | 314.33 | 313.54 | 300.75 | 295.04 |
| *K*(0 K) ($J/m^3$) | **$2.12 \times 10^6$** | $2.08 \times 10^6$ | $1.90 \times 10^6$ | $2.04 \times 10^6$ | $1.41 \times 10^6$ |
| *K*(297 K) ($J/m^3$) | $1.23 \times 10^5$ | $1.12 \times 10^5$ | $1.03 \times 10^5$ | $2.66 \times 10^4$ | $1 \times 10^4$ |

Atomistic spin dynamics simulations of the Curie temperature of $Cr_{1-x}Nb_xTe_2$ are performed using VAMPIRE package[83] utilizing the adaptive Monte-Carlo integrator[84] with temperature-dependent magnetization dynamics[85] using the Heisenberg Hamiltonian shown in section 2.4 of the main text. The exchange parameters $J_{||}$, $J_{\perp}$, and anisotropy constant *K* are extracted from density functional theory (DFT) calculations and reported in Figure 6. The simulations were carried out using a time step of 1 fs, with 100,000 equilibration steps followed by 100,000 steps for averaging the magnetization at each temperature. The temperature increased in increments of 5 K. Periodic boundary conditions were applied along all three directions using a supercell of size $15a \times 15\sqrt{3}a \times 15c$.

The temperature-dependent magnetization results are shown in Figure S4.1c and $T_C$ values vs the Nb doping factor $x$ are displayed in Table S4.2. The results indicate a systematic decrease in $T_C$ with increasing the Nb concentration. This trend is primarily attributed to the suppression of the IP exchange interaction $J_{\parallel}$ (Fig. 6d).

**Table S4.2.** Deduced $T_C$ of $Cr_{1-x}Nb_xTe_2$ vs Nb doping factor $x$.

| $x$ | 0 | 0.05 | 0.1 | 0.15 | 0.2 |
|---|---|---|---|---|---|
| $T_C$ (K) | 269 | 246 | 213 | 161 | 85 |

The Monte Carlo derived $T_C$ values are consistently lower than the measured values, which can be attributed to limitations of the modeling. First, the virtual crystal approximation may lose accuracy at higher Nb doping levels (e.g., at 20%) due to the local chemical disorder. Second, the spin Hamiltonian includes only nearest neighbor IP and OOP exchange interactions and neglect longer interactions, which may underestimate the overall exchange strength which hence lead to a lower predicted $T_C$.

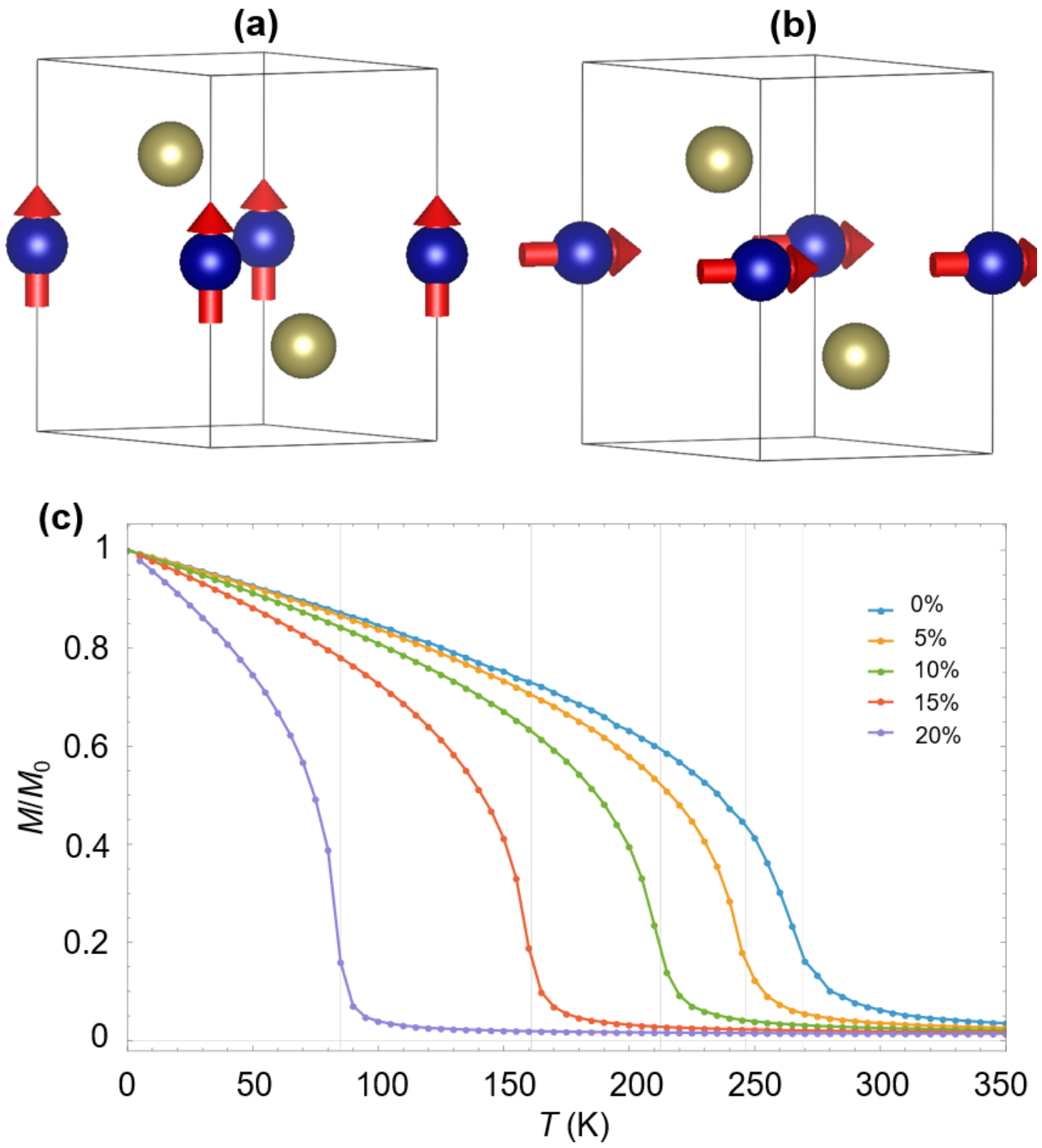

**Figure S4.1.** (a,b) Spin configurations of $Cr_{1-x}Nb_xTe_2$ ($x$ = 0 – 0.2) used in the magnetic anisotropy calculations: (a) OOP and (b) IP spin orientations. (c) Atomic scale simulation of magnetization as a function of temperature using the Monte Carlo method.

To further elucidate the microscopic origin of the anisotropy evolution, we calculated the electronic band structures and density of states (DOS) of $Cr_{1-x}Nb_xTe_2$. As shown in Figures S4.2(d-e), Nb substitution modifies the electronic structure. In particular, bands near the k point shift upward, indicating reduced electron occupation, while bands around the Γ point shift slightly downward. These changes enhance the density of states near the Fermi level.

Within second-order perturbation theory, the magnetocrystalline anisotropy $E_{\mathrm{MCA}}$ can be expressed as:

$$E_{\mathrm{MCA}} = \frac{\xi^2}{4}\sum_{o,u}\frac{|\langle\psi_o|L_z|\psi_u\rangle|^2 - |\langle\psi_o|L_x|\psi_u\rangle|^2}{\varepsilon_u - \varepsilon_o},$$

where $\psi_o$ and $\psi_u$ denote occupied and unoccupied states with energies $\varepsilon_o$ and $\varepsilon_u$, respectively, and $\xi$ is the spin orbit coupling (SOC) constant.[86] In this system, the dominant contributions arise from the transition metal $d$ orbitals. The orbital resolved projected DOS in Figures S4.2(a-c) show that the occupied spin majority and unoccupied spin minority $d_{xy}$ and $d_{x^2-y^2}$ orbitals shift toward the Fermi level upon Nb doping, enhancing SOC matrix elements mediated by the $L_z$ operator $\left(\frac{|\langle\psi_o|L_z|\psi_u\rangle|^2}{\varepsilon_u-\varepsilon_o}\right)$ due to reduced denominator $\varepsilon_u - \varepsilon_o$. In contrast, the $L_x$ related channels including $d_{z^2} \leftrightarrow d_{zy}$ and $d_{zx} \leftrightarrow d_{x^2-y^2}/d_{xy}$ exhibit weaker enhancement due to minimal DOS changes in the $d_{zx}/d_{zy}$ states (Figure S4.2c) near the Fermi level. As a result, the increase in the $L_z$ contribution outweighs that in the $L_x$ contribution. Consequently, the anisotroy $E_{\mathrm{MAE}} \propto |L_z|^2 - |L_x|^2$ remains negative but its magnitude decreases, indicating a weakened in-plane magnetic anisotropy.

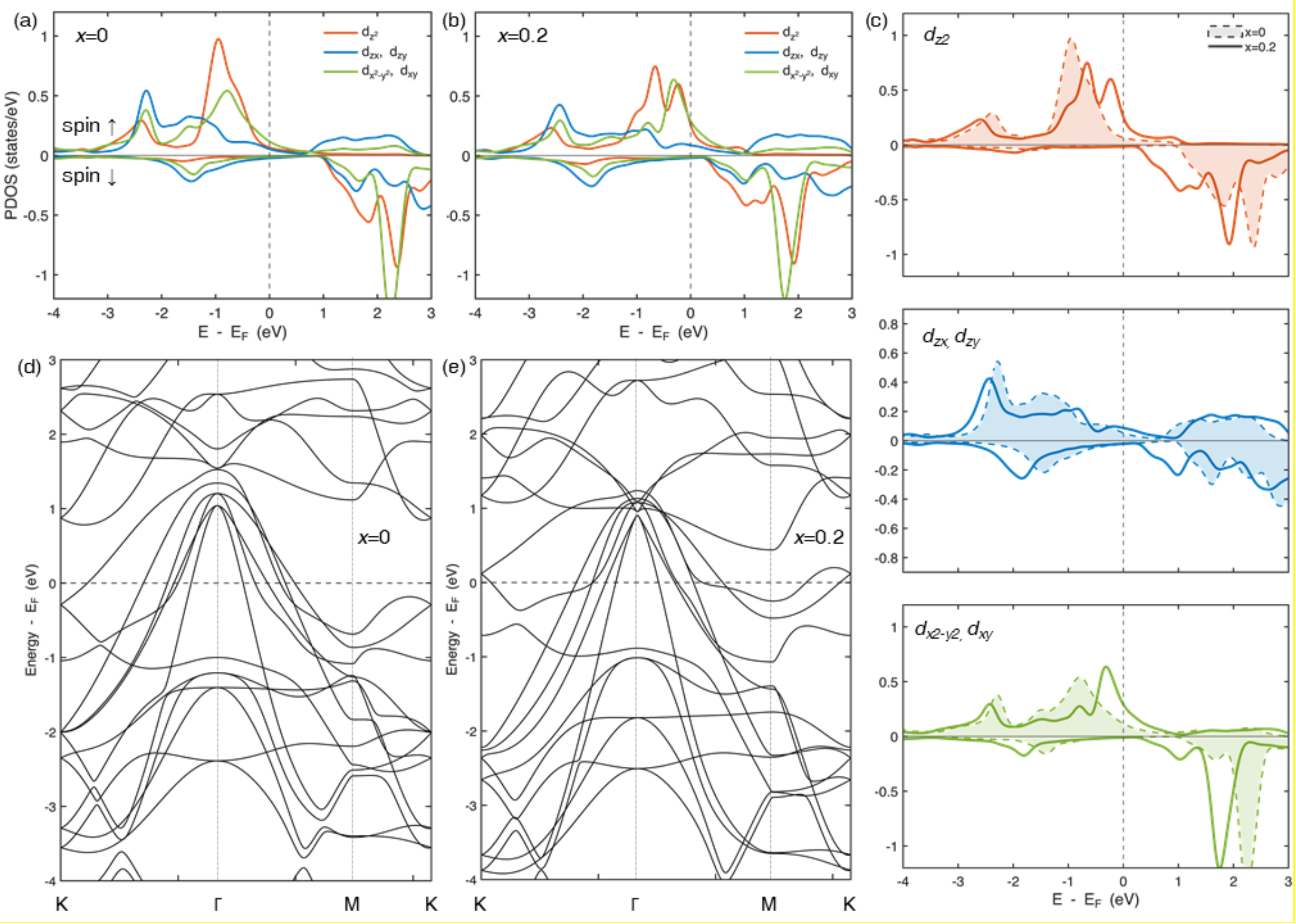

**Figure S4.2**. (a,b) Orbital resolved projected DOS for $x = 0$ and $x = 0.2$, respectively. The contributions from $d_{z^2}$, $d_{zx}/d_{zy}$, and $d_{x^2-y^2}/d_{xy}$ orbitals are shown. Positive and negative DOS correspond to spin up and spin down channels, respectively. (c) Direct comparison of projected DOS between $x = 0$ and $x = 0.2$. (d,e) Corresponding band structures for $x = 0$ and $x = 0.2$.